\documentclass{emulateapj}   

\newcommand{\kms}{{\hbox {km\thinspace s$^{-1}$}}}
\newcommand{\Lsun}{{\hbox {L$_\odot$}}}
\newcommand{\Msun}{{\hbox {M$_\odot$}}}

\newcommand{\cmt}{{\hbox {cm$^{-3}$}}}
\newcommand{\cmd}{{\hbox {cm$^{-2}$}}}

\newcommand{\hdo}{{\hbox {H$_{2}$O}}}

\newcommand{\tgas}{{\hbox {$T_{\mathrm{gas}}$}}}

\newcommand{\tdust}{{\hbox {$T_{\mathrm{dust}}$}}}

\shorttitle{High-lying OH absorption, [C {\sc ii}] deficits, and
extreme $L_{\mathrm{FIR}}/M_{\mathrm{H2}}$ ratios in galaxies}
\shortauthors{Gonz\'alez-Alfonso et al.}

\begin{document}


\title{High-lying OH absorption, [C {\sc ii}] deficits, and
extreme $L_{\mathrm{FIR}}/M_{\mathrm{H2}}$ ratios in galaxies}


\author{E. Gonz\'alez-Alfonso\altaffilmark{1}, J. Fischer\altaffilmark{2},
  E. Sturm\altaffilmark{3}, J. Graci\'a-Carpio\altaffilmark{3}, 
  S. Veilleux\altaffilmark{4}, M. Mel\'endez\altaffilmark{4}, 
  D. Lutz\altaffilmark{3}, A. Poglitsch\altaffilmark{3}, 
  S. Aalto\altaffilmark{5}, N. Falstad\altaffilmark{5}, 
  H. W. W. Spoon\altaffilmark{6}, D. Farrah\altaffilmark{7}, 
  A. Blasco\altaffilmark{1}, C. Henkel\altaffilmark{8,9},
  A. Contursi\altaffilmark{3}, A. Verma\altaffilmark{10}, 
  M. Spaans\altaffilmark{11}, 
  H. A. Smith\altaffilmark{12}, M. L. N. Ashby\altaffilmark{12},
  S. Hailey-Dunsheath\altaffilmark{13},
  S. Garc\'{\i}a-Burillo\altaffilmark{14}, 
  J. Mart\'{\i}n-Pintado\altaffilmark{15}, 
  P. van der Werf\altaffilmark{16}, R. Meijerink\altaffilmark{16},  
  R. Genzel\altaffilmark{3}
}
\affil{$^1$Universidad de Alcal\'a, Departamento de F\'{\i}sica
     y Matem\'aticas, Campus Universitario, E-28871 Alcal\'a de Henares,
     Madrid, Spain}
\affil{$^2$Naval Research Laboratory, Remote Sensing Division, 4555
     Overlook Ave SW, Washington, DC 20375, USA}
\affil{$^3$Max-Planck-Institute for Extraterrestrial Physics (MPE),
  Giessenbachstra{\ss}e 1, 85748 Garching, Germany}
\affil{$^4$Department of Astronomy, University of Maryland, College Park, MD
  20742, USA} 
\affil{$^5$Department of Earth and Space Sciences, Chalmers University of
  Technology, Onsala Space Observatory, Onsala, Sweden } 
\affil{$^6$Cornell University, Astronomy Department, Ithaca, NY 14853, USA} 
\affil{$^7$Department of Physics, Virginia Tech, Blacksburg, VA 24061, USA} 
\affil{$^8$Max-Planck-Institut f\"ur Radioastronomie, Auf dem H\"ugel 69,
  53121, Bonn, Germany}
\affil{$^9$Astronomy Department, Kind Abdulaziz University,
  P.O. Box 80203, Jeddah 21589, Saudi Arabia}
\affil{$^{10}$University of Oxford, Oxford Astrophysics, Denys Wilkinson
  Building, Keble Road, Oxford, OX1 3RH, UK}
\affil{$^{11}$Kapteyn Astronomical Institute, University of Groningen, PO Box
  800, 9700 AV Groningen, The Netherlands}  
\affil{$^{12}$Harvard-Smithsonian Center for Astrophysics, 60 Garden Street,
  Cambridge, MA 02138, USA} 
\affil{$^{13}$California Institute of Technology, 1200 E. California
Blvd., Pasadena, CA 91125, USA}
\affil{$^{14}$Observatorio Astron\'omico Nacional (OAN)-Observatorio de
  Madrid, Alfonso XII 3, 28014, Madrid, Spain}  
\affil{$^{15}$CSIC/INTA, Ctra de Torrej\'on a Ajalvir, km 4, 28850, Torrej\'on
  de Ardoz, Madrid, Spain}  
\affil{$^{16}$Sterrewacht Leiden, Leiden University, PO Box 9513, 2300 RA,
  Leiden, The Netherlands}  



\begin{abstract}
{\it Herschel}/PACS observations of 29 local (Ultra-)Luminous
Infrared Galaxies, including both starburst and AGN-dominated 
sources as diagnosed in the mid-infrared/optical, show that the
equivalent width of the absorbing OH 65 $\mu$m $\Pi_{3/2}$ $J=9/2-7/2$ line
($W_{\mathrm{eq}}(\mathrm{OH65})$) with lower level energy 
$E_{\mathrm{low}}\approx300$ K, is anticorrelated with the [C {\sc ii}]158
$\mu$m line to far-infrared luminosity ratio, and correlated with
the far-infrared luminosity per unit gas mass and with the
60-to-100 $\mu$m far-infrared color. 
While all sources are in the active $L_{\mathrm{IR}}/M_{\mathrm{H2}}>50$
\Lsun/\Msun\ mode as derived from previous CO line studies, the OH65
absorption shows a bimodal distribution with a discontinuity at
$L_{\mathrm{FIR}}/M_{\mathrm{H2}}\approx100$ \Lsun/\Msun.
In the most buried sources, OH65 probes material partially responsible for the
silicate $9.7$ $\mu$m absorption.
Combined with observations of the OH 71 $\mu$m $\Pi_{1/2}$ $J=7/2-5/2$ doublet
($E_{\mathrm{low}}\approx415$ K),
radiative transfer models characterized by the equivalent dust temperature,
\tdust, and the continuum optical depth at 100 $\mu$m, $\tau_{100}$, 
indicate that strong [C {\sc ii}]158 $\mu$m deficits are
associated with far-IR thick 
($\tau_{100}\gtrsim0.7$, $N_{\mathrm{H}}\gtrsim10^{24}$ cm$^{-2}$), warm
($T_{\mathrm{dust}}\gtrsim60$ K) 
structures where the OH 65 $\mu$m absorption is produced, most
likely in circumnuclear disks/tori/cocoons. 
With their high $L_{\mathrm{FIR}}/M_{\mathrm{H2}}$ ratios and columns, 
the presence of these structures is expected to give rise to strong 
[C {\sc ii}] deficits.
$W_{\mathrm{eq}}(\mathrm{OH65})$ probes the
fraction of infrared luminosity arising from these compact/warm
environments, which is $\gtrsim30-50$\% in sources with high
$W_{\mathrm{eq}}(\mathrm{OH65})$. 
Sources with high $W_{\mathrm{eq}}(\mathrm{OH65})$ have surface densities 
of both $L_{\mathrm{IR}}$ and $M_{\mathrm{H2}}$ higher than inferred from the
half-light (CO or UV/optical) radius, tracing coherent structures that 
represent the most buried/active stage of (circum)nuclear starburst-AGN
co-evolution.  
\end{abstract}

\keywords{galaxies: ISM --- galaxies: evolution --- infrared: galaxies --- line: formation }

\section{Introduction} \label{intro}

From the first spectroscopic observations of (ultra)luminous infrared galaxies
((U)LIRGs) in the far-infrared (far-IR) domain with the Infrared Space
Observatory (ISO), evidence was found that the strength of fine-structure
lines (from both ions and atoms) in emission are generally anticorrelated with
the depth and excitation of the molecular lines observed in absorption
\citep{fis99}. The most commonly observed line, the fine-structure 
[C {\sc ii}]157.7 $\mu$m transition (hereafter [C {\sc ii}]), tends to exhibit
a strong deficit with respect to the far-IR luminosity in ULIRGs relative to
less luminous systems \citep{luh98,luh03}. In normal galaxies, the 
[C {\sc ii}]/FIR luminosity ratio remains nearly constant ($0.1-1$\%), while
it decreases in galaxies with warmer far-IR colors \citep{mal01,dia13}. On the
other hand, studies of individual templates (Arp~220 and Mrk~231) indicated
that high far-IR radiation densities associated with the nuclear regions of
galaxies with [C {\sc ii}] deficits, are required to account for the
observed high-lying molecular absorption \citep[][hereafter G-A08]{gon04,gon08}.

\begin{figure*}
\includegraphics[angle=0,scale=.60]{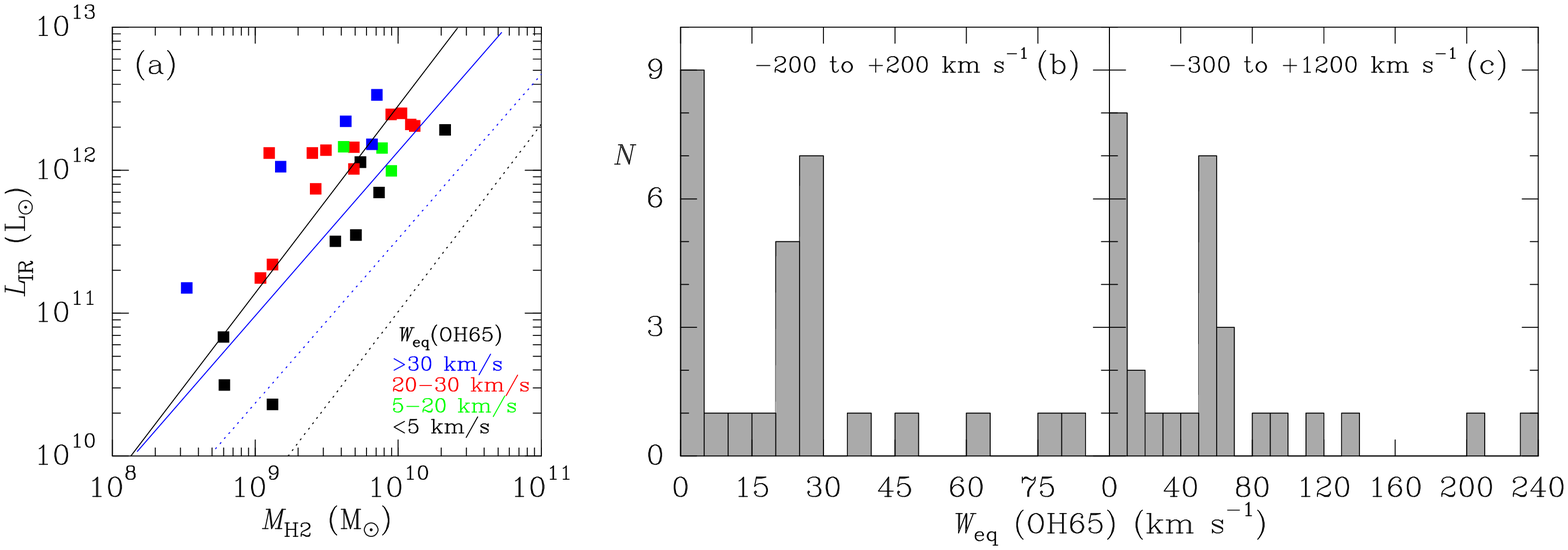}
\caption{a) Infrared luminosities as a function of the H$_2$ masses for
  our sample of local (U)LIRGs. Symbol colors indicate values of
    $W_{\mathrm{eq}}(\mathrm{OH65})$ in four bins. The solid and dotted
    lines are fits to ULIRG-SMG-QSO mergers and spiral-BzK-normal galaxies,
    respectively, 
  given by \cite{dad10} (black) and \cite{gen10} (blue). 
  b) Histogram of $W_{\mathrm{eq}}(\mathrm{OH65})$ calculated between
  $-200$ and $+200$ \kms\ in bins of 5 \kms, showing an apparent bimodal
distribution with a long tail extending up to 85 \kms. 
 c) Histogram of $W_{\mathrm{eq}}(\mathrm{OH65})$ calculated
  between $-300$ and $+1200$ \kms\ (covering essentially
  the whole doublet), with bins of 10 \kms. }    
\label{distrib}
\end{figure*}

The launch of the {\it Herschel} Space Observatory \citep{pil10}
has dramatically improved the sensitivity of these measurements. 
Observations with the PACS spectrometer \citep{pog10} soon revealed that
the observed deficit of [C {\sc ii}] relative to the far-IR emission applies
to all far-IR fine-structure lines 
\citep[][hereafter G-C11]{fis10,far13,gra11}.
G-C11 also showed that the deficits are better correlated with
$L_{\mathrm{FIR}}/M_{\mathrm{gas}}$ than with $L_{\mathrm{FIR}}$,
while PACS observations of three (U)LIRGs
with strong line deficits, NGC~4418, Arp~220, and Mrk~231, showed
deep absorption in high-lying molecular lines \citep[][hereafter
G-A12 and G-A14]{gon12,gon14}. NGC~4418 is a case in point, as it shows the
highest [C {\sc ii}] deficit, a moderate $L_{\mathrm{IR}}\sim1.5\times10^{11}$
\Lsun\ but a high $L_{\mathrm{FIR}}/M_{\mathrm{gas}}\approx400$ \Lsun/\Msun\
(G-C11), and the highest-lying \hdo\ absorption among 
all galaxies with full FIR spectra (G-A12).

To explore the connection between intense far-IR fields and both the
highly excited molecular gas and the [C {\sc ii}] deficit, we investigate
the relationship between the OH $^2\Pi_{3/2}\,J=9/2-7/2$ transition at
$65.2$ $\mu$m (hereafter OH65) with $E_{\mathrm{low}}\approx300$ K, and the 
[C {\sc ii}] line, $L_{\mathrm{FIR}}/M_{\mathrm{gas}}$, the $9.7$ $\mu$m silicate
absorption, and the far-IR colors, also using measurements of
the OH $^2\Pi_{1/2}\,J=7/2-5/2$ transition at $71.2$ $\mu$m (hereafter OH71,
$E_{\mathrm{low}}\approx415$ K) in galaxies for which it is available.
OH is a versatile molecule with high abundances in active regions 
including photodissociated regions (PDRs), cosmic-ray dominated regions
(CRDRs), and X-ray dominated regions (XDRs)
\citep[e.g.][]{goi02,goi11,mei11,gon13}, and traces powerful galactic-scale
molecular outflows in some sources 
\citep[][hereafter V13; G-A14]{fis10,stu11,spo13,vei13}
mostly associated with large AGN luminosity fractions and luminosities.
In extragalactic sources, the OH65 doublet (when detected) is
  absorption-dominated, indicating that 
  the excitation of the lower $^2\Pi_{3/2}\,J=7/2$ level is governed by
  radiative (rather than collisional) processes\footnote{
   Collisional excitation of $^2\Pi_{3/2}\,J=7/2$ followed by OH65 absorption 
   is not dominant owing to the high $A-$Einstein coefficient of the 84
   $\mu$m ($^2\Pi_{3/2}\,7/2-5/2$) transition; efficient OH65 absorption
   involves a high radiation density such that it will also dominate the 
    excitation of $^2\Pi_{3/2}\,J=7/2$ under reasonable physical
    condictions.}. The OH65 pumping thus 
  involves successive absorptions in the 119, 84, 
  and finally in the 65 $\mu$m doublet with high $A-$Einstein coefficients
  ($0.14$, $0.51$, and $1.2$ s$^{-1}$, see the energy level diagram of OH in 
  G-A14), thus ensuring an excellent probe of strong far-IR fields.

\begin{figure}
\includegraphics[angle=0,scale=.43]{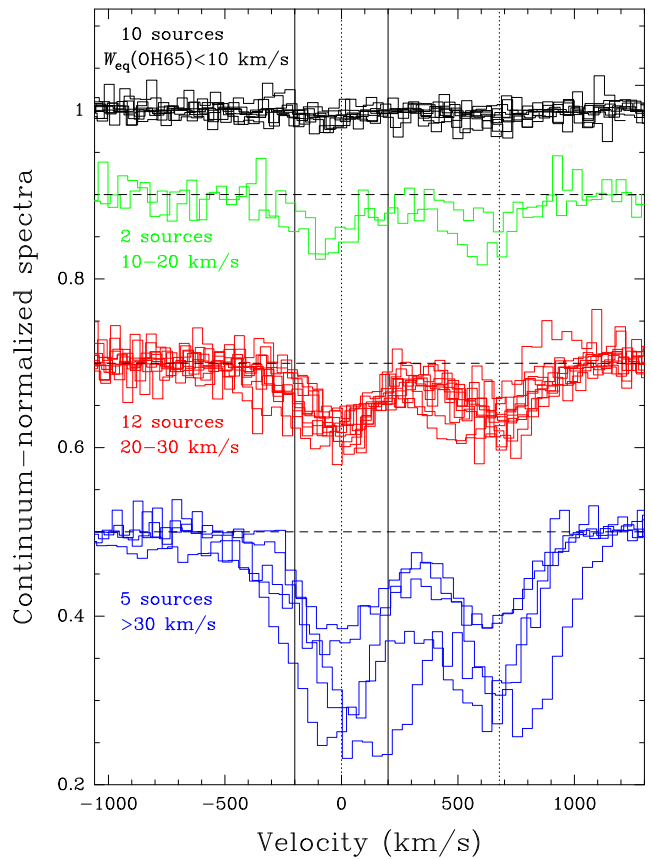}
\caption{The OH $^2\Pi_{3/2}$ 65 $\mu$m $J=\frac{9}{2}-\frac{7}{2}$
  continuum-normalized spectra in all galaxies in the sample, with the
  velocity plotted relative to the rest-frame wavelength of the blue component
  of the doublet ($J=\frac{9}{2}^--\frac{7}{2}^+$ line at $65.1316$
  $\mu$m). The spectra are grouped according to the values of the equivalent
  width measured between $-200$ and $+200$ \kms\ around the blue
  component (indicated by the solid vertical lines and listed in
  Table~\ref{tbl-2}). The dotted vertical lines
  indicate the positions of the two components of the doublet. The green, red,
  and blue spectra are vertically shifted for clarity.} 
\label{spec65}
\end{figure}

\begin{figure}
\includegraphics[angle=0,scale=.43]{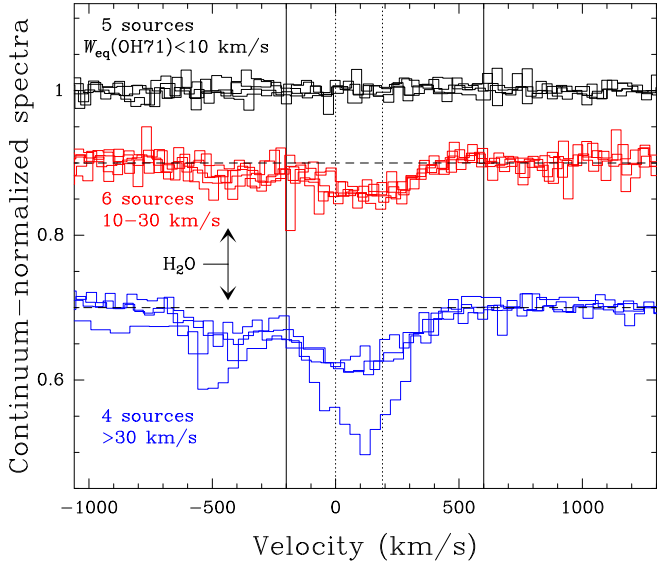}
\caption{The OH $^2\Pi_{1/2}$ 71 $\mu$m $J=\frac{7}{2}-\frac{5}{2}$
  continuum-normalized spectra in all 15 galaxies for which it is available,
  with the velocity plotted relative to the rest-frame wavelength of the blue
  component of the doublet at $71.171$ $\mu$m. The dotted vertical lines
  indicate the positions of the two $\Lambda$-components of the doublet, which
  are blended into a single spectral feature. The spectra are grouped
  according to the values of the equivalent width measured between $-200$ and
  $+600$ \kms\ (indicated by the vertical solid lines and listed in
  Table~\ref{tbl-2}).  The red and blue 
  spectra are vertically shifted for clarity. The position of the
  \hdo\ $5_{24}-4_{13}$ line ($E_{\mathrm{low}}\approx400$ K) is indicated.}
\label{spec71}
\end{figure}

\begin{figure*}
\includegraphics[angle=0,scale=.75]{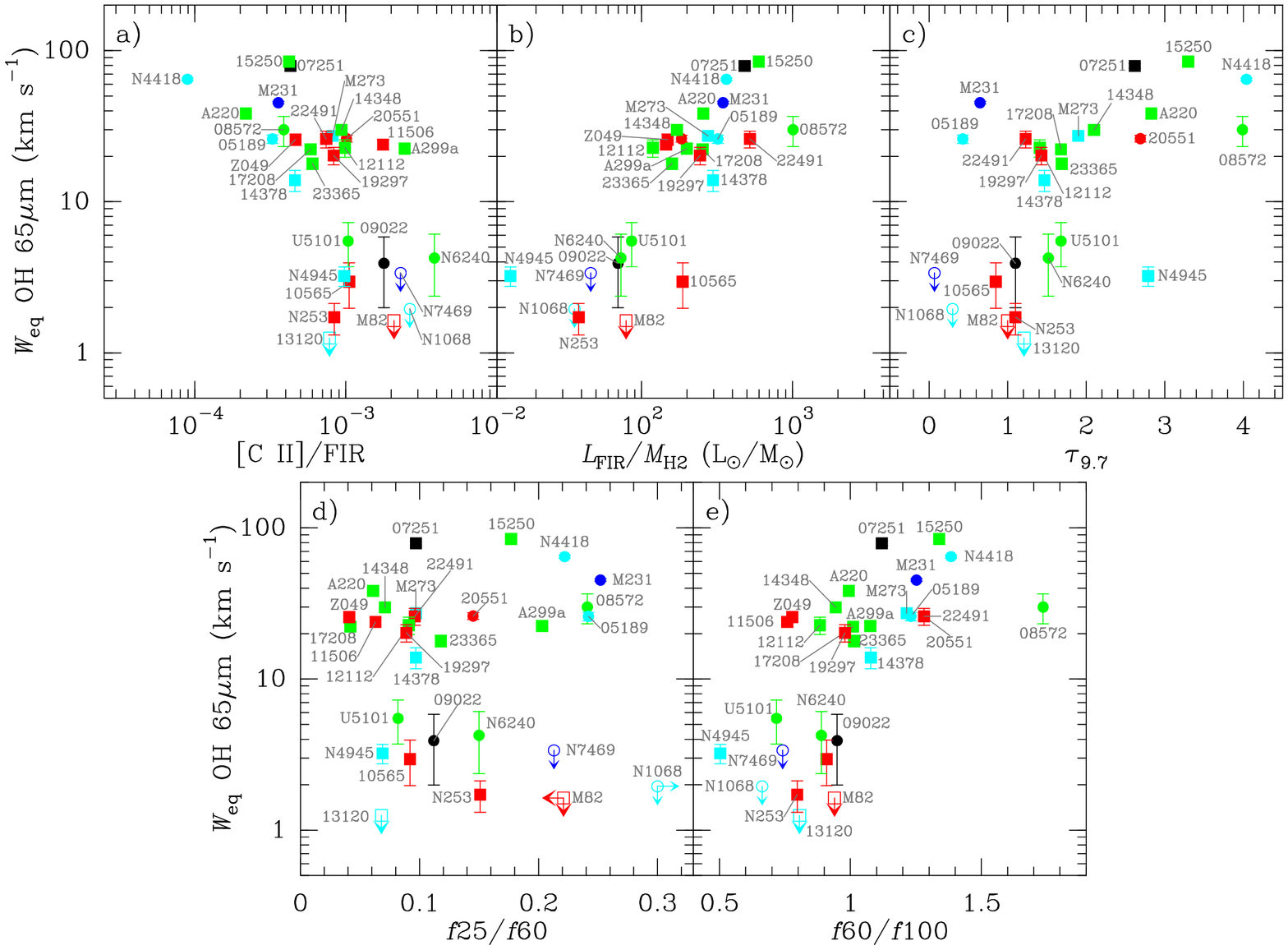}
\caption{Equivalent width of the OH $\Pi_{3/2}$
  $J=\frac{9}{2}^--\frac{7}{2}^+$ line at 65.132 $\mu$m (the blue
  component of the 65 $\mu$m doublet) between
  $-200$ and $+200$ \kms, as a function of (a) the [C {\sc
    ii}]158 $\mu$m line to FIR ratio, (b) the far-IR
  luminosity per unit gas mass, (c) the apparent optical depth of
    the silicate absorption at 9.7 $\mu$m \citep[from][]{spo07}, 
  (d) the 25-to-60 $\mu$m color, 
  and (e) the 60-to-100 $\mu$m color. Abbreviated source names are
  indicated. Red, green, blue, light-blue, and black colors indicate 
  H{\sc ii}, LINER, Seyfert-1, Seyfert-2, and unclassified optical spectral
  types, respectively \citep[from][V09, or
  NED/SIMBAD]{vei95,vei99,ver06,rup05,gar06,kim98}. Circles and 
  squares indicate sources with fractional AGN contribution to the
  bolometric luminosity of $\alpha_{\mathrm{AGN}}\ge50$\% and $<50$\%,
  respectively, as derived from $f15/f30$ (V09). Pearson $\chi^2$
  independence-tests give chance probabilities 
  $P=0.012-0.0-0.043-0.74-0.025$ for panels a-b-c-d-e, respectively.
} 
\label{correl}
\end{figure*}

\begin{figure}
\includegraphics[angle=0,scale=.70]{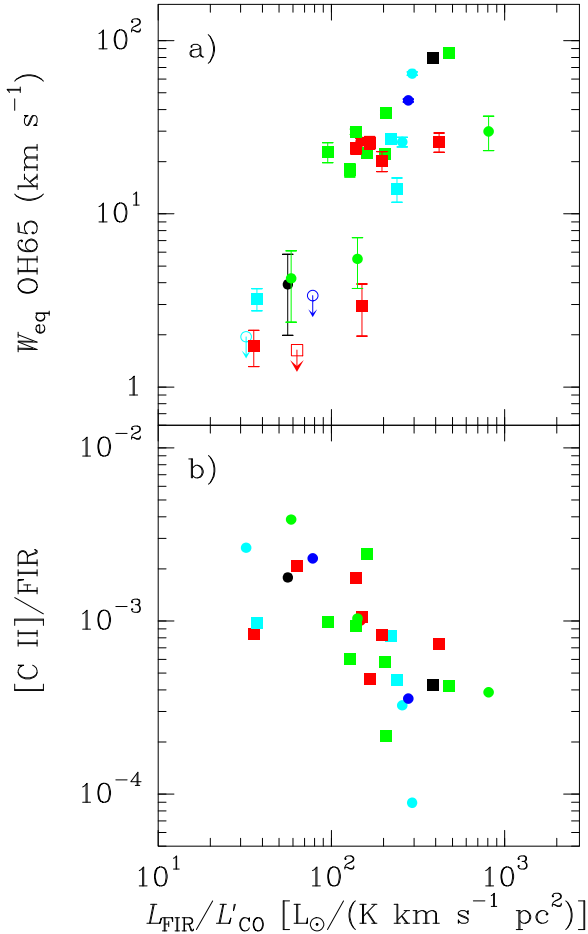}
\caption{a) Equivalent width of the OH65 transition between
  $-200$ and $+200$ \kms\ around the blue component of the doublet, and (b)
  the [C {\sc ii}]158 $\mu$m line to FIR ratio, as a function of the the
  far-IR to CO (1-0) luminosity ratio in our galaxy sample. Symbol colors and
  shapes have the same meaning as in Fig.~\ref{correl}.}
\label{correl2}
\end{figure}

\section{Observations} \label{obs}

We have used {\it Herschel}/PACS observations of the OH65
transition in local ($z<0.1$) galaxies, included in the {\it Herschel}
guaranteed time key program SHINING (PI: E. Sturm) and in three 
OT programs (PIs: E. Gonz\'alez-Alfonso; J. Fischer, S. Hailey-Dunsheath). 
Table~\ref{tbl-1} lists the sample galaxies and their properties.
With a total of 29 galaxies, the sample is biased towards ULIRGs 
\citep[including 17 out of the 18 most luminous sources in the IRAS
  Revised Bright Galaxy Sample;][]{san03}, 
but also contains less luminous systems including Seyferts and H{\sc ii}
galaxies. The OH71 doublet was observed in a subsample of 15 sources.
The locations of the targets in the $L_{\mathrm{IR}}-M_{\mathrm{H2}}$ plane,
shown in Fig.~\ref{distrib}a, indicate that all except NGC~4945 (with
  $L_{\mathrm{IR}}=2.3\times10^{10}$ \Lsun) belong to
the high $L_{\mathrm{IR}}/M_{\mathrm{H2}}$ mode ($>50$ \Lsun/\Msun) as
compared with normal/disk galaxies \citep{dad10,gen10}. 

The data were reduced using the standard PACS reduction and
calibration pipeline included in HIPE $6.0$ and $10.0$, 
recalibrating the data with a reference telescope spectrum obtained from
  observations of Neptune (G-C11). A few spectra were also reduced using
  HIPE $12.0$. There are moderate calibration differences (typically
  $\approx10$\% and in a few sources up to $\approx20$\%)
in both line and continuum flux densities between HIPE $6.0$ and
$10.0-12.0$, but the continuum-normalized spectra used to measure the OH
equivalent widths were found essentially identical in the different
versions. Likewise, the PACS-based ratios presented below ([C {\sc ii}]/FIR
and $f60/f100$) are not sensitive to global calibration issues.

The OH65 spectra, displayed in Fig.~\ref{spec65}, 
are dominated by absorption at central velocities, but some sources show 
detection of blue wings (e.g. Mrk~231, G-A14). Redshifted reemission by the
blue component of the doublet from outflowing gas on the 
far side of the nucleus has the effect of decreasing the relative strength of
the red component of the doublet (see asymmetrical doublets in
Fig.~\ref{spec65}). 
We measured the equivalent width ($W_{\mathrm{eq}}$) of the OH $\Pi_{3/2}$
$J=\frac{9}{2}-\frac{7}{2}$ doublet between $-200$ and $+200$ \kms\ around the
blue component of the doublet, and between $-300$ and $+1200$ \kms, covering
essentially the whole doublet (Table~\ref{tbl-2} and
Fig.~\ref{spec65}). This work is focussed on 
the structures traced by the excited OH at central velocities, so hereafter we
primarily study $W_{\mathrm{eq}}(\mathrm{OH65})$ in the $-200$ and $+200$
\kms\ velocity range while the outflowing gas component will be treated in a
separate study. The OH71 spectra, shown in Fig.~\ref{spec71}, also indicate
peak absorption at central velocities.


Most sources in the sample are unresolved with the PACS
$9"\times9"$ spatial resolution, and thus $W_{\mathrm{eq}}(\mathrm{OH65})$
measured from the central PACS spaxel applies to the whole galaxy. 
For resolved sources (M82, Arp~299a, NGC~1068, NGC~253, and NGC~4945),
$W_{\mathrm{eq}}(\mathrm{OH65})$ was measured from the 25-spaxel
combined spectra, covering a field of view (FoV) of $47"\times 47"$.
Likewise, $W_{\mathrm{eq}}(\mathrm{OH71})$, the flux of the 
[C {\sc ii}] line, the FIR, and the flux densities at 60 and 100
$\mu$m ($f60$ and $f100$) were all integrated over the total PACS FoV.
Even for the extended sources, the PACS $f60$ and $f100$ agree to within
$20$\% with the 60 and 100 $\mu$m {\it IRAS} flux densities
\citep{san03,sur04}, indicating that PACS recovers the bulk of the
galaxy far-IR continuum emission and the $W_{\mathrm{eq}}(\mathrm{OH})$
represent global values. The {\it IRAS} 25 $\mu$m flux densities ($f25$) are
then also used in our analysis. 
All PACS-measured [C {\sc ii}] line fluxes agree with the {\it ISO}-LWS
values or upper limits \citep{bra08} within 40\%, and most of them within 25\%.
  Most values of $M_{\mathrm{H2}}$ were estimated from the
  spatially-integrated CO(1-0) luminosities from previous studies
  (Table~\ref{tbl-1}) by using a conversion factor 
$\alpha_{\mathrm{CO}}$ decreasing 
with $f60/f100$ (G-C11). Since the sources in our sample are warm,
$\alpha_{\mathrm{CO}}=0.8$ was mostly applied (Table~\ref{tbl-1}); only
UGC~5101, NGC~7469, and especially NGC~4945 have 
significantly higher $\alpha_{\mathrm{CO}}>1.2$.

\begin{deluxetable*}{lcccccccccl}
\tabletypesize{\scriptsize}
\tablecaption{Sample galaxies}
\tablewidth{0pt}
\tablehead{
\colhead{Galaxy} & 
\colhead{$D$} &
\colhead{$L_{\mathrm{IR}}$} &
\colhead{$L'_{\mathrm{CO}}$} &
\colhead{$\alpha_{\mathrm{CO}}$} &
\colhead{$M_{\mathrm{H2}}$} & 
\colhead{$L_{\mathrm{FIR}}/M_{\mathrm{H2}}$} &
\colhead{$\tau_{9.7}$} & 
\colhead{$f25/f60$} &
\colhead{$f60/f100$} &
\colhead{R$_{\mathrm{CO}}$} \\
\colhead{name} & 
\colhead{(Mpc)} & 
\colhead{($10^{11}$ \Lsun)} &
\colhead{($10^{9}$ L$_{\mathrm{l}}$)} & 
\colhead{(\Msun/L$_{\mathrm{l}}$)} & 
\colhead{($10^{9}$ \Msun)} & 
\colhead{($10^{2}$ \Lsun/\Msun)} &
\colhead{} &
\colhead{} &
\colhead{} &
\colhead{} \\
\colhead{(1)} & 
\colhead{(2)} & 
\colhead{(3)} & 
\colhead{(4)} & 
\colhead{(5)} &
\colhead{(6)} & 
\colhead{(7)} & 
\colhead{(8)} &
\colhead{(9)} & 
\colhead{(10)} &
\colhead{(11)} 
}
\startdata
 IRAS 07251-0248  &  $398.0 $ & $22.00 $ & $ 5.35 $ & $0.803 $ & $ 4.30 $ & $ 4.82 $ & $2.62 $ & $0.10 $ & $1.12 $ & Sp                     \\                  
 IRAS 09022-3615  &  $268.0 $ & $19.20 $ & $26.56 $ & $0.803 $ & $21.33 $ & $ 0.70 $ & $1.10 $ & $0.11 $ & $0.95 $ & GC                     \\                  
 M 82             &  $  3.9 $ & $ 0.68 $ & $ 0.75 $ & $0.803 $ & $ 0.60 $ & $ 0.79 $ & $1.00 $ & $0.22 $ & $0.94 $ & We05\tablenotemark{a}  \\                  
 IRAS 13120-5453  &  $136.0 $ & $18.60 $ &          &          &          &          & $1.21 $ & $0.07 $ & $0.81 $ &                        \\                  
 NGC 253          &  $  3.3 $ & $ 0.31 $ & $ 0.66 $ & $0.929 $ & $ 0.61 $ & $ 0.38 $ & $1.10 $ & $0.15 $ & $0.80 $ & Ho97\tablenotemark{b}  \\                  
 NGC 1068         &  $ 18.0 $ & $ 3.18 $ & $ 4.06 $ & $0.896 $ & $ 3.64 $ & $ 0.36 $ & $0.30 $ & $0.47 $ & $0.66 $ & Sc83\tablenotemark{c}  \\                  
 IRAS F05189-2524 &  $186.0 $ & $13.80 $ & $ 3.90 $ & $0.803 $ & $ 3.13 $ & $ 3.20 $ & $0.43 $ & $0.24 $ & $1.23 $ & Pa12                   \\                  
 IRAS F08572+3915 &  $261.0 $ & $13.20 $ & $ 1.56 $ & $0.803 $ & $ 1.25 $ & $10.06 $ & $3.99 $ & $0.24 $ & $1.74 $ & So97\tablenotemark{d}  \\                  
 UGC 5101         &  $173.0 $ & $ 9.91 $ & $ 5.44 $ & $1.650 $ & $ 8.97 $ & $ 0.86 $ & $1.68 $ & $0.08 $ & $0.72 $ & So97\tablenotemark{d}  \\                  
 IRAS F10565+2448 &  $193.0 $ & $11.40 $ & $ 6.79 $ & $0.803 $ & $ 5.45 $ & $ 1.87 $ & $0.85 $ & $0.09 $ & $0.91 $ & So97\tablenotemark{d}  \\                  
 Arp 299a         &  $ 46.5 $ & $ 7.41 $ & $ 3.30 $ & $0.803 $ & $ 2.65 $ & $ 1.99 $ &         & $0.20 $ & $1.08 $ & Ca99                   \\                  
 IRAS F11506-3851 &  $ 49.0 $ & $ 2.19 $ & $ 1.39 $ & $0.954 $ & $ 1.32 $ & $ 1.45 $ &         & $0.06 $ & $0.76 $ & Mi90\tablenotemark{e}  \\                  
 IRAS F12112+0305 &  $333.0 $ & $20.40 $ & $16.25 $ & $0.803 $ & $13.06 $ & $ 1.19 $ & $1.41 $ & $0.09 $ & $0.88 $ & Ch09                   \\                  
 NGC 4418         &  $ 35.9 $ & $ 1.50 $ & $ 0.41 $ & $0.803 $ & $ 0.33 $ & $ 3.65 $ & $4.04 $ & $0.22 $ & $1.38 $ & Pa12                   \\                  
 Mrk 231          &  $186.0 $ & $33.70 $ & $ 8.84 $ & $0.803 $ & $ 7.10 $ & $ 3.46 $ & $0.65 $ & $0.25 $ & $1.25 $ & So97\tablenotemark{d}  \\                  
 NGC 4945         &  $  3.4 $ & $ 0.23 $ & $ 0.48 $ & $2.754 $ & $ 1.32 $ & $ 0.14 $ & $2.79 $ & $0.07 $ & $0.50 $ & He94\tablenotemark{f}  \\                  
 Mrk 273          &  $166.0 $ & $14.50 $ & $ 6.11 $ & $0.803 $ & $ 4.91 $ & $ 2.75 $ & $1.90 $ & $0.10 $ & $1.22 $ & So97\tablenotemark{d}  \\                  
 IRAS F14348-1447 &  $376.0 $ & $20.90 $ & $15.23 $ & $0.803 $ & $12.24 $ & $ 1.72 $ & $2.10 $ & $0.07 $ & $0.94 $ & Ch09                   \\                  
 IRAS F14378-3651 &  $304.0 $ & $14.60 $ & $ 5.18 $ & $0.803 $ & $ 4.16 $ & $ 2.98 $ & $1.47 $ & $0.10 $ & $1.08 $ & Mi90\tablenotemark{e}  \\                  
 Zw 049.057       &  $ 58.0 $ & $ 1.76 $ & $ 0.97 $ & $1.116 $ & $ 1.09 $ & $ 1.49 $ &         & $0.04 $ & $0.78 $ & Pa12                   \\                  
 IRAS F15250+3609 &  $245.0 $ & $10.60 $ & $ 1.88 $ & $0.803 $ & $ 1.51 $ & $ 5.97 $ & $3.30 $ & $0.18 $ & $1.34 $ & Ch09                   \\                  
 Arp 220          &  $ 79.4 $ & $15.20 $ & $ 8.17 $ & $0.803 $ & $ 6.56 $ & $ 2.56 $ & $2.83 $ & $0.06 $ & $0.99 $ & So97\tablenotemark{d}  \\                  
 NGC 6240         &  $106.0 $ & $ 6.98 $ & $ 9.16 $ & $0.803 $ & $ 7.36 $ & $ 0.73 $ & $1.52 $ & $0.15 $ & $0.89 $ & So97\tablenotemark{d}  \\                  
 IRAS F17207-0014 &  $187.0 $ & $25.00 $ & $13.12 $ & $0.803 $ & $10.54 $ & $ 2.53 $ & $1.68 $ & $0.04 $ & $1.01 $ & Pa12                   \\                  
 IRAS F19297-0406 &  $383.0 $ & $24.60 $ & $11.13 $ & $0.803 $ & $ 8.94 $ & $ 2.44 $ & $1.43 $ & $0.09 $ & $0.98 $ & So97\tablenotemark{d}  \\                  
 IRAS F20551-4250 &  $185.0 $ & $10.20 $ & $ 6.10 $ & $0.803 $ & $ 4.90 $ & $ 1.84 $ & $2.69 $ & $0.14 $ & $1.28 $ & Mi90\tablenotemark{e}  \\                  
 IRAS F22491-1808 &  $343.0 $ & $13.20 $ & $ 3.12 $ & $0.803 $ & $ 2.51 $ & $ 5.22 $ & $1.23 $ & $0.10 $ & $1.28 $ & Ch09                   \\                  
 NGC 7469         &  $ 64.5 $ & $ 3.52 $ & $ 2.98 $ & $1.697 $ & $ 5.06 $ & $ 0.46 $ & $0.07 $ & $0.21 $ & $0.74 $ & Pa12                   \\                  
 IRAS F23365+3604 &  $281.0 $ & $14.30 $ & $ 9.61 $ & $0.803 $ & $ 7.72 $ & $ 1.59 $ & $1.69 $ & $0.12 $ & $1.01 $ & So97\tablenotemark{d} 
\enddata
\label{tbl-1}
\tablecomments{(1) Galaxy name; 
  (2) Distance to the galaxy; adopting a flat Universe with $H_0=71$ km
   s$^{-1}$ Mpc$^{-1}$ and $\Omega_{\mathrm{M}}=0.27$. For
   some nearby galaxies alternative distances are used;  
  (3) IR luminosity ($8-1000$ $\mu$m), estimated using the fluxes in the four
  {\it IRAS} bands \citep{san03,sur04}; 
  (4) CO (1-0) luminosity from previous studies (col 12), 
  L$_{\mathrm{l}}=\mathrm{K\,km\,s^{-1}\, pc^2}$; 
  (5) Conversion factor, $M_{\mathrm{H2}}=\alpha_{\mathrm{CO}}\times
  L'_{\mathrm{CO}}$, $\alpha_{\mathrm{CO}}$ increases with decreasing $f60/f100$; 
  (6) H$_2$ mass;
  (7) Far-IR luminosity ($40-500$ $\mu$m) to H$_2$ mass ratio;
  (8) Apparent optical depth of the silicate absorption at 9.7 $\mu$m
  \citep{spo07};  
  (9) Continuum $25$-to-$60$ $\mu$m flux density ratio;
  (10) Continuum $60$-to-$100$ $\mu$m flux density ratio;
  (11) Reference for $L'_{\mathrm{CO}}$. 
      Sp: Spoon, unpublished data taken with the IRAM 30m telescope; 
      GC: Graci\'a-Carpio, unpublished CO(2-1) data taken with APEX, and
      assuming $L'_{\mathrm{CO2-1}}/L'_{\mathrm{CO1-0}}=0.7$;
      We05: \cite{wei05}; Ho97: \cite{hou97}; Sc83: \cite{sco83}; Pa12:
      \cite{pap12}; So97: \cite{sol97}; Ca99: \cite{cas99}; Mi90:
      \cite{mir90}; Ch09: \cite{chu09}; He94: \cite{hen94}.
}
\tablenotetext{a}{CO (1-0) flux within the central $3\times3$ kpc$^2$, which
  reduces to half within the inner $1\times1$ kpc$^2$.}
\tablenotetext{b}{Corrected for our adopted distance.}
\tablenotetext{c}{Adopting $S_{\nu}/T_{\mathrm{A}}^*=42$ Jy/K for FCRAO
  \citep{pap12} and correcting for extended emission.}
\tablenotetext{d}{CO (1-0) fluxes corrected for $S_{\nu}/T_{\mathrm{MB}}=4.95$
  Jy/K, IRAM-30m telescope.}
\tablenotetext{e}{Adopting $S_{\nu}/T_{\mathrm{R}}^*=24.5$ Jy/K for SEST.}
\tablenotetext{f}{Corrected for extended emission, a CO (1-0) flux of
  $1.7\times10^4$ Jy \kms\ is estimated.} 
\end{deluxetable*}

\section{Results} \label{res}

$W_{\mathrm{eq}}(\mathrm{OH65})$ shows a bimodal distribution
(Fig.~\ref{distrib}b-c and Fig.~\ref{spec65})
with peaks at $<5$ and $20-30$ \kms\ and a long tail extending up 
to 85 \kms. A Pearson $\chi^2$-test comparing the observed
distribution for central velocities
in Fig.~\ref{distrib}b with a flat distribution gives a
$P-$value $\approx0.01$, which remains low ($0.03$) when the full doublet
  is considered  (Fig.~\ref{distrib}c). 
While most ULIRGs are strong in OH65, some of them (IRAS~09022-3615, 
UGC~5101, IRAS~F10565+2448) are weak and, conversely, there are three 
sources with moderate luminosities that are strong in OH65 (NGC~4418, 
Zw~049.057, and IRAS~F11506-3851). Nevertheless,
sources with high OH65 absorption have, on average, higher infrared
luminosities for {\em fixed} $M_{\mathrm{H2}}$ (Fig.~\ref{distrib}a). 
Among galaxies with low $W_{\mathrm{eq}}(\mathrm{OH65})<10$
  \kms, some still have clear detections of OH65 (IRAS~F10565+2448,
  UGC~5101, IRAS~09022-3615, NGC~4945, NGC~253), but no trace of OH65
  absorption is found in others (M~82, NGC~1068, NGC~7469).
OH71 is detected in the 10 sources with 
$W_{\mathrm{eq}}(\mathrm{OH65})>20$ \kms\ (Fig.~\ref{spec71},
Table~\ref{tbl-2}) and is  undetected in the remaining five sources where
OH65 is weak or undetected.

Figure~\ref{correl}a shows that $W_{\mathrm{eq}}(\mathrm{OH65})$ and 
[C {\sc ii}]/FIR are anticorrelated, with
data points mainly concentrated in the 
upper-left and lower-right quadrants. 
The [C {\sc ii}]/FIR value that separates the two regimes, $\approx10^{-3}$,
is similar to that found by \cite{luh03} for the onset of the [C {\sc ii}]
deficit. Within the limitations of our sample, the data in Fig.~\ref{correl}a
show {\em that galaxies with a strong [C {\sc ii}] deficit
  have relatively deep OH65 absorption}. 
Around the critical value of [C {\sc ii}]/FIR$\sim10^{-3}$, however, 
there are galaxies with high and low OH65 absorption. The
objects with high $W_{\mathrm{eq}}(\mathrm{OH65})$ and also high
[C {\sc ii}]/FIR are Arp~299 and IRAS~11506-3851, the former having a complex
structure with multiple nuclei\footnote{The FoV
  was centered in the Arp~299 $A1-$nucleus \citep[IC~694,][]{aal97} where the
  peak OH65 absorption is found, but includes extended [C {\sc ii}] and CO
  emitting regions.} and the latter showing the coldest $f60/f100$ color
  among sources with $W_{\mathrm{eq}}(\mathrm{OH65})>10$ \kms.
On the other hand, IRAS~13120-5453, NGC~253, IRAS~10565+2448, NGC~4945,
  and UGC~5101 are examples of galaxies at or close to the onset of the
  [C {\sc ii}] deficit that lack high OH65 absorption. 

Since $L_{\mathrm{FIR}}/M_{\mathrm{H2}}$ and [C {\sc ii}]/FIR are
anticorrelated (G-C11), a positive correlation between the former and
$W_{\mathrm{eq}}(\mathrm{OH65})$ is anticipated, and this is seen
in Fig.~\ref{correl}b. A discontinuity in OH65 absorption is found at
$L_{\mathrm{FIR}}/M_{\mathrm{H2}}\sim100$ \Lsun/\Msun, similar to the one
that separates mergers from normal galaxies 
in the [C {\sc ii}]/FIR-$L_{\mathrm{FIR}}/M_{\mathrm{gas}}$ plane
(G-C11).\footnote{We note that M~82, with a global
    $L_{\mathrm{FIR}}/M_{\mathrm{H2}}\sim80$ 
  \Lsun/\Msun, has a CO(1-0) luminosity of half the total value within the
  PACS FoV \citep[][see also Table~\ref{tbl-1}]{wei05}, and hence
  $L_{\mathrm{FIR}}/M_{\mathrm{H2}}\sim160$ \Lsun/\Msun\ in this
  region. Nevertheless, inhomogeneities are also expected in unresolved
  sources (\S\ref{compc} below), and thus we use the global values of
  $M_{\mathrm{H2}}$ in Fig.~\ref{correl}b.} It is worth noting that
  $M_{\mathrm{H2}}$ is the most uncertain value in Fig.~\ref{correl}; however,
  Fig.~\ref{correl2} shows that the correlation remains when using 
  the CO (1-0) luminosity directly, and that [C {\sc ii}]/FIR shows a marked
  anticorrelation with $L_{\mathrm{FIR}}/L'_{\mathrm{CO}}$. 

Figure~\ref{correl}c relates $W_{\mathrm{eq}}(\mathrm{OH65})$ and 
the apparent optical depth of the 9.7 $\mu$m silicate 
feature \citep{spo07}.
With the exception of the nearly edge-on galaxy NGC~4945, sources with
weak $W_{\mathrm{eq}}(\mathrm{OH65})<10$ \kms\ have moderate
$\tau_{9.7}\lesssim1.7$, suggesting that a 
significant fraction of the silicate absorption in the most obscured 
($\tau_{9.7}\gtrsim2$) objects is produced by the material responsible for the
OH65 absorption. This is consistent with the finding by \cite{gou12} 
that the dominant contribution to the silicate feature in Compton-thick AGNs
is dust located in the host galaxy, which we identify in part with the OH65
structure\footnote{We consider the OH65 region to be part of the host even
    though it is circumnuclear (see footnote~\ref{foot:size}), because it is
    much larger than the pc-scale tori surrounding Compton-thick AGNs. In
    Arp~220 and NGC~4418, where the [O {\sc i}] 63 $\mu$m line is observed
    primarily in absorption, a significant fraction of the Si absorption is
    likely produced in extended regions (G-A12).}.  
The trends seen in Figs.~\ref{correl}a and c are also 
consistent with the observed anti-correlation between [C {\sc ii}]/FIR and the
strength of the 9.7 $\mu$m silicate feature \citep{dia13}. 
With increasing silicate obscuration, the low-lying OH 119 $\mu$m
doublet shows increasing absorption relative to emission, suggesting 
obscuration of the OH re-emission behind a far-IR optically thick
(circum)nuclear region \citep[][V13]{spo13}; the data suggest that this is 
also accompanied by high OH excitation as measured by the
  high-lying OH65 doublet. High OH65 absorption, however, does not guarantee
deep silicate absorption, as seen for a number of sources including the
  Seyferts Mrk~231 and IRAS~F05189-2524.

Figure~\ref{correl}d-e shows no apparent correlation between
$W_{\mathrm{eq}}(\mathrm{OH65})$ and $f25/f60$, but shows a
positive correlation with the $f60/f100$ far-IR color. 
In light of the anti-correlation between $W_{\mathrm{eq}}(\mathrm{OH65})$
  and [C {\sc ii}]/FIR (Fig.~\ref{correl}a), this is consistent with  
the decline observed in [C {\sc ii}]/FIR with warmer far-IR colors
\citep{mal01,luh03,dia13}. While $f60/f100>1$ is 
associated with the structure traced by the OH65 absorption, there is a
vertical overlap of sources showing strong and weak OH65 absorption
within the $f60/f100\lesssim1$ bin. 
Finally, no relationship is found between
$W_{\mathrm{eq}}(\mathrm{OH65})$ and either the optical spectral type or 
the fractional contribution of the AGN to the bolometric luminosity as
diagnosed from $f15/f30$ \citep[method 6 in][hereafter V09]{vei09}.

\begin{figure*}
\begin{center}
\includegraphics[angle=0,scale=.75]{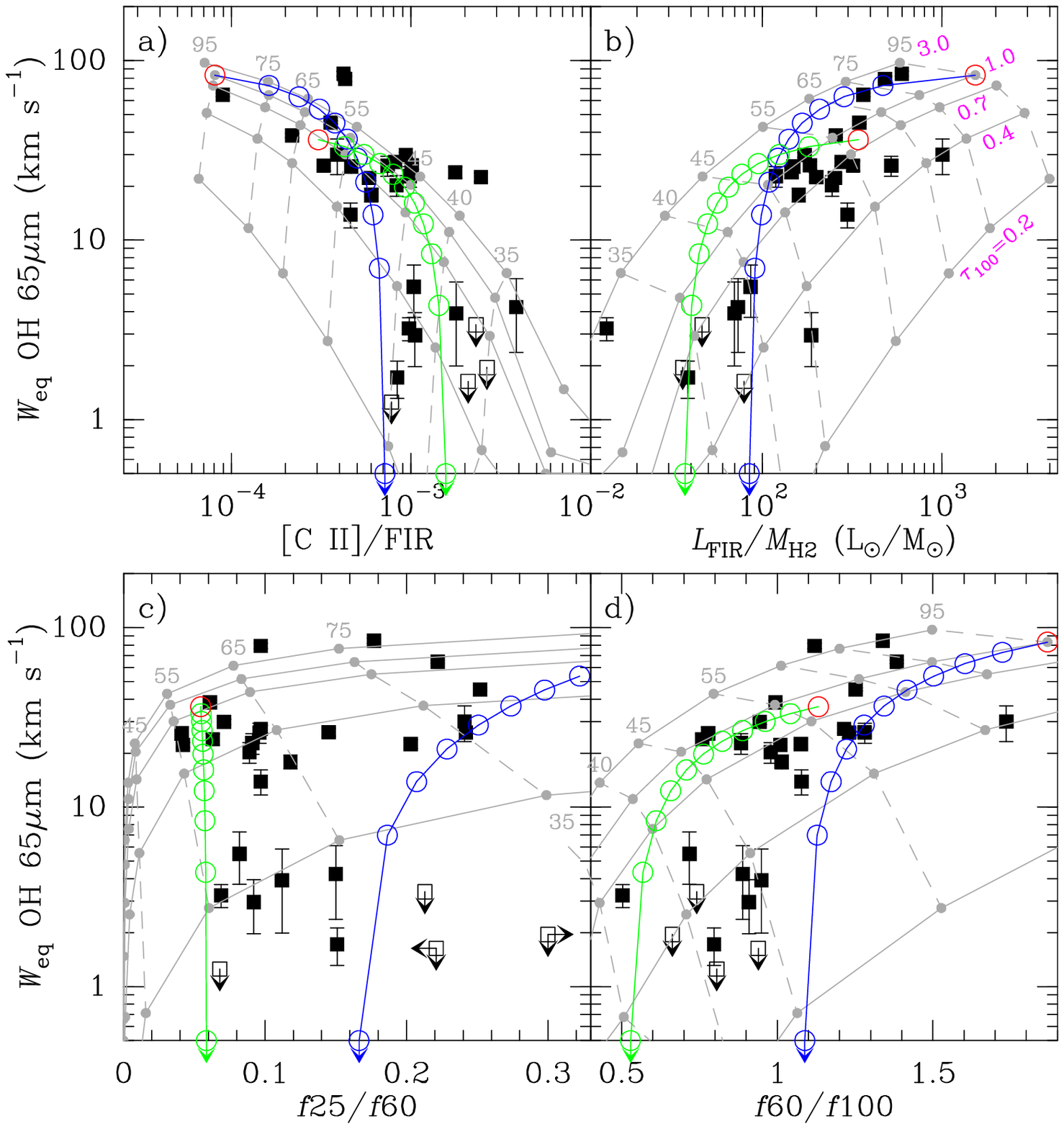}
\end{center}
\caption{Same as Fig.~\ref{correl} with results from single-component
  (\S\ref{singlec}, grey symbols and lines) and composite (\S\ref{compc}, 
  blue and green) radiative
  transfer models overlaid. The single-component models are characterized by
  the dust temperature, \tdust, and the continuum optical depth at 100 $\mu$m,
  $\tau_{100}$. Solid grey lines connect results 
  with constant $\tau_{100}$ (indicated in magenta in panel b),
  and dashed grey lines connect results with constant \tdust\
  (indicated with grey numbers, in K). The composite models have 
  three components: two optically
  thin components with $T_{\mathrm{dust}}=65$ and 30 K, generating the
  bulk of the [C {\sc ii}] 158 $\mu$m emission 
  but negligible OH 65 $\mu$m absorption, and one optically
  thick ($\tau_{100}=1$) and very warm ($T_{\mathrm{dust}}=95$ K for model
  $M_1$, blue symbols, and $T_{\mathrm{dust}}=60$ K for model $M_2$, green
  symbols) component, responsible for the OH 65 $\mu$m absorption. 
  The mix of components is described by the parameters
  $\beta_{65-30}=L_{\mathrm{IR}}^{65}/(L_{\mathrm{IR}}^{65}+L_{\mathrm{IR}}^{30})$ and
$\beta_{\mathrm{thick}}=L_{\mathrm{IR}}^{\mathrm{thick}}/L_{\mathrm{IR}}^{\mathrm{total}}$
(see \S\ref{compc}). 
  $\beta_{65-30}=0.75$ and $0.25$ for models $M_1$ and $M_2$,
  respectively. Along the sequence for both blue and green lines,
  $\beta_{\mathrm{thick}}$ is varied from 0 to 1 by intervals of $0.1$, with
  $\beta_{\mathrm{thick}}=1$ indicated with a red circle.
}   
\label{models}
\end{figure*}

\begin{deluxetable*}{llllcccc}
\tabletypesize{\scriptsize}
\tablecaption{{\it Herschel}/PACS measurements of 
$W_{\mathrm{eq}}(\mathrm{OH65})$, $W_{\mathrm{eq}}(\mathrm{OH71})$, 
and [C {\sc ii}]/FIR}
\tablewidth{0pt}
\tablehead{
\colhead{Galaxy} & 
\colhead{$W_{\mathrm{eq}}(\mathrm{OH65})$} & 
\colhead{$W_{\mathrm{eq}}(\mathrm{OH65})$} & 
\colhead{$W_{\mathrm{eq}}(\mathrm{OH71})$} &
\colhead{[C {\sc ii}]/FIR} &
\colhead{OBSID}  &
\colhead{OBSID}  &
\colhead{OBSID}  \\
\colhead{name} & 
\colhead{(\kms)} & 
\colhead{(\kms)} & 
\colhead{(\kms)} & 
\colhead{($10^{-3}$)} &
\colhead{OH65} &
\colhead{OH71} &
\colhead{[C {\sc ii}]158 $\mu$m} \\
\colhead{(1)} & 
\colhead{(2)} & 
\colhead{(3)} & 
\colhead{(4)} &
\colhead{(5)} &
\colhead{(6)} &
\colhead{(7)} &
\colhead{(8)}
}
\startdata
 IRAS 07251-0248  & $79.2(2.7)$ & $200.9( 5.4)$ &               &  $ 0.43 $ & $1342207824$ &              & $1342207825$  \\
 IRAS 09022-3615  & $ 3.9(1.9)$ & $ 15.3( 3.8)$ &               &  $ 1.79 $ & $1342209406$ &              & $1342209403$  \\
 M 82             & $<1.6     $ & $< 3.5      $ & $< 2.4      $ &  $ 2.08 $ & $1342186966$ & $1342186966$ & $1342186798$  \\
 IRAS 13120-5453  & $<1.3     $ & $< 2.5      $ & $< 2.5      $ &  $ 0.78 $ & $1342214630$ & $1342248349$ & $1342214629$  \\
 NGC 253          & $ 1.7(0.4)$ & $  3.5( 0.8)$ &               &  $ 0.84 $ & $1342237602$ &              & $1342199415$  \\
 NGC 1068         & $<2.0     $ & $< 3.7      $ & $< 2.6      $ &  $ 2.65 $ & $1342203128$ & $1342203120$ & $1342191154$  \\
 IRAS F05189-2524 & $26.0(1.7)$ & $ 63.1( 3.3)$ & $ 15.6( 3.3)$ &  $ 0.33 $ & $1342219445$ & $1342248557$ & $1342219442$  \\
 IRAS F08572+3915 & $29.9(6.8)$ & $ 38.2(12.6)$ & $ 23.5( 3.4)$ &  $ 0.39 $ & $1342208954$ & $1342245405$ & $1342208952$  \\
 UGC 5101         & $ 5.5(1.8)$ & $ 12.7( 3.5)$ &               &  $ 1.03 $ & $1342208950$ &              & $1342208949$  \\
 IRAS F10565+2448 & $ 3.0(1.0)$ & $  6.6( 1.9)$ &               &  $ 1.05 $ & $1342207790$ &              & $1342207788$  \\
 Arp 299a         & $22.4(1.3)$ & $ 54.3( 2.5)$ & $ 16.4( 2.8)$ &  $ 2.45 $ & $1342254241$ & $1342254242$ & $1342208906$  \\
 IRAS F11506-3851 & $23.9(1.7)$ & $ 56.8( 3.4)$ & $ 12.7( 3.2)$ &  $ 1.77 $ & $1342248551$ & $1342248549$ & $1342248549$  \\
 IRAS F12112+0305 & $22.7(3.0)$ & $ 54.9( 6.0)$ &               &  $ 0.99 $ & $1342210833$ &              & $1342210832$  \\
 NGC 4418         & $64.6(1.4)$ & $136.1( 2.6)$ & $ 68.8( 1.3)$ &  $ 0.09 $ & $1342202115$ & $1342202107$ & $1342210830$  \\
 Mrk 231          & $45.2(0.9)$ & $111.7( 1.7)$ & $ 35.8( 2.8)$ &  $ 0.36 $ & $1342207782$ & $1342253534$ & $1342186811$  \\
 NGC 4945         & $ 3.2(0.5)$ & $  7.5( 0.9)$ &               &  $ 0.98 $ & $1342247790$ &              & $1342212221$  \\
 Mrk 273          & $27.3(1.3)$ & $ 81.8( 2.5)$ & $ 37.1( 2.1)$ &  $ 0.82 $ & $1342207803$ & $1342257292$ & $1342207802$  \\
 IRAS F14348-1447 & $29.9(1.0)$ & $ 69.9( 2.0)$ &               &  $ 0.94 $ & $1342224244$ &              & $1342224242$  \\
 IRAS F14378-3651 & $13.9(2.2)$ & $ 22.5( 4.4)$ &               &  $ 0.46 $ & $1342204339$ &              & $1342204338$  \\
 Zw 049.057       & $25.8(2.2)$ & $ 63.5( 4.2)$ & $ 16.9( 2.3)$ &  $ 0.46 $ & $1342248366$ & $1342248365$ & $1342248365$  \\
 IRAS F15250+3609 & $84.5(2.7)$ & $238.5( 5.0)$ &               &  $ 0.42 $ & $1342213754$ &              & $1342211825$  \\
 Arp 220          & $38.4(1.0)$ & $ 93.4( 2.0)$ & $ 40.5( 1.5)$ &  $ 0.22 $ & $1342238937$ & $1342238928$ & $1342191306$  \\
 NGC 6240         & $ 4.2(1.9)$ & $< 6.8      $ & $< 5.2      $ &  $ 3.86 $ & $1342216624$ & $1342251372$ & $1342216623$  \\
 IRAS F17207-0014 & $22.2(0.7)$ & $ 55.0( 1.4)$ & $ 17.4( 2.2)$ &  $ 0.58 $ & $1342229694$ & $1342252279$ & $1342229693$  \\
 IRAS F19297-0406 & $20.2(2.7)$ & $ 45.4( 5.1)$ &               &  $ 0.83 $ & $1342208893$ &              & $1342208891$  \\
 IRAS F20551-4250 & $26.1(1.3)$ & $ 56.1( 2.6)$ &               &  $ 1.01 $ & $1342208936$ &              & $1342208934$  \\
 IRAS F22491-1808 & $26.0(3.3)$ & $ 53.9( 6.2)$ &               &  $ 0.74 $ & $1342211826$ &              & $1342211825$  \\
 NGC 7469         & $<3.4     $ & $< 6.6      $ & $< 6.7      $ &  $ 2.30 $ & $1342235674$ & $1342235841$ & $1342211171$  \\
 IRAS F23365+3604 & $17.9(1.6)$ & $ 50.6( 3.1)$ &               &  $ 0.60 $ & $1342212517$ &              & $1342212515$
 \enddata
\label{tbl-2}
\tablecomments{(1) Galaxy name; 
  (2) Equivalent width of OH65 between $-200$ and $+200$ \kms\ around the blue
  component of the doublet, covering the central velocity component;  
  (3) Equivalent width of OH65 between $-300$ and $+1200$ \kms, around the blue
  component of the doublet, covering the whole doublet;
  (4) Equivalent width of OH71 between $-200$ and $+600$ \kms, covering the
  whole doublet. Numbers in parenthesis indicate $1\sigma$
  uncertainties and upper limits are $2\sigma$. These $\sigma$ values are
  calculated as $\sigma_{\mathrm{rms}}^{\mathrm{norm}}\times\Delta V/n^{1/2}$, where
  $\sigma_{\mathrm{rms}}^{\mathrm{norm}}$ is the RMS noise of the
  continuum-normalized spectrum and $\Delta V$ and $n$ are the velocity
  coverage and number of channels over which $W_{\mathrm{eq}}(\mathrm{OH65})$
  is measured;
  (5) [C {\sc ii}] to FIR \citep[$42.5-122.5$ $\mu$m, following ][]{hel88}
  flux ratio; 
  (6) Identification number for the OH65 observations;
  (7) Identification number for the OH71 observations;
  (8) Identification number for the [C {\sc ii}] 158 $\mu$m observations.
}
\end{deluxetable*}

\section{Radiative transfer models} \label{rt}

To characterize the overall physical conditions derived from the present
observations, and to interpret the trends shown in Fig.~\ref{correl}, 
phenomenological radiative transfer models have been generated
(G-A14 and references therein). The model sources are spherical and assume
uniform physical conditions, parameterized by
the dust temperature (\tdust), the continuum optical depth at 100 $\mu$m
($\tau_{100}$), the gas temperature and density (\tgas\ and
$n_{\mathrm{H}}$), the OH and C$^+$ column densities ($N_{\mathrm{OH}}$ and 
$N_{\mathrm{C^+}}$), and the velocity dispersion ($\Delta V$).

\subsection{Single-component models} \label{singlec}

Initially, we naively assume that the OH65 absorption and [C {\sc ii}]
emission arise from the same region and that the covering factor of the
continuum by the excited OH is unity; these assumptions represent only a first
approach to the interpretation of the observations but still enable us to
extract some general conclusions. 
To decrease the number of free parameters, we approximate
some of them according to previous chemical or radiative transfer models: 
$(i)$ The gas column density ($N_{\mathrm{H}}$) is directly related to
$\tau_{100}$ by adopting a standard gas-to-dust ratio by mass of 100 and
a dust mass opacity coefficient at 100 $\mu$m of 
$\kappa_{100}=44.5$ cm$^2$ g$^{-1}$:
$N_{\mathrm{H}}=1.3\times10^{24}\,\tau_{100}\,\mathrm{cm^{-2}}$ 
\citep{gon14b}. 
$(ii)$ $N_{\mathrm{OH}}$ is fixed by assuming an OH abundance relative to H of
$2.5\times10^{-6}$ (G-A12, G-A14). 
$(iii)$ We assume that the [C {\sc ii}] emission is dominated by PDRs
\citep{far13} and $N_{\mathrm{C^+}}$ is estimated on the basis of previous models
\citep[e.g.][G-C11]{abe09,kau99}. For a single PDR, 
$N_{\mathrm{C^+}}\sim10^{18}$ \cmd\ is typically inferred for high 
incident far-UV radiation intensity $G_0$, and
$T_{\mathrm{dust}}\sim10\times G_0^{0.2}$ K characterizes the warm dust in
PDRs \citep{hol91}. We set $X_{\mathrm{C^+}}=(2-6)\times10^{-5}$ for
$T_{\mathrm{dust}}=30-90$ K to approximately account for these results and
calculate $N_{\mathrm{C^+}}$ based on $N_{\mathrm{H}}(\tau_{100})$. 
$(iv)$ We adopt $n_{\mathrm{H}}=5\times10^4$ \cmt\ and $T_{\mathrm{gas}}=150$ K,
i.e. high density conditions appropriate for the circumnuclear regions of
(U)LIRGs, ensuring that the [C {\sc ii}] transition is thermalized and
  emits at nearly the maximum emission per C$^+$ ion \citep{tie85}. 
The OH65 transition is pumped through absorption of far-IR photons and it is
thus not sensitive to $n_{\mathrm{H}}$ and \tgas\ (G-A08). 
$(v)$ $\Delta V=100$ \kms\ in all models, describing the velocity
  dispersion along a characteristic line of sight. This has no effect in case
of optically thin lines. As we show below, however, high columns and thus many
overlapping (shadowing) regions characterize the environments where the OH65
absorption is produced, most likely forming a medium bound to the total
  potential of the galaxy center \citep{dow93,sol97}. 
High $\Delta V$ is thus used to simulate the velocity dispersion
due to random cloud-cloud motions, non-rigid rotation, and high-scale
turbulence.

The above prescription leaves only two free parameters, \tdust\ and
$\tau_{100}$, the properties of the far-IR emission from
which the observables $W_{\mathrm{eq}}(\mathrm{OH65})$, 
[C {\sc ii}]/FIR, $L_{\mathrm{FIR}}/M_{\mathrm{H2}}$, $f25/f60$, and
$f60/f100$ are calculated.  
Figure~\ref{models} overlays the single-component model results (gray symbols
and lines) and the data. Dashed gray lines connect model results for
fixed \tdust, and solid gray lines connect model results for given
$\tau_{100}$. While the [C {\sc ii}] deficit is only a function of \tdust\ 
(dashed lines are nearly vertical in Fig.~\ref{models}a),
$W_{\mathrm{eq}}(\mathrm{OH65})$ is sensitive 
to both \tdust\ and $\tau_{100}$. Since our adopted OH and C$^+$ abundances
are probably upper limits, Fig.~\ref{models}a indicates that moderately high
$T_{\mathrm{dust}}\gtrsim50$ K and high column densities
($\tau_{100}\gtrsim0.5$ or
$N_{\mathrm{H}}\gtrsim6\times10^{23}\,\mathrm{cm^{-2}}$) are the lower limits
that characterize objects with $W_{\mathrm{eq}}(\mathrm{OH65})\gtrsim20$
\kms. This is consistent with model 
results in Fig.~\ref{models}b, where the $L_{\mathrm{FIR}}/M_{\mathrm{H2}}$
values for these sources are reproduced with similar physical conditions.

For low $\tau_{100}$, $W_{\mathrm{eq}}(\mathrm{OH65})$ increases
  supralinearly with $\tau_{100}$ (Fig.~\ref{models}) as a result of both the
  increasing strength of the radiation field that pumps the lower level of the
  transition, and the increasing OH column. 
However, the OH65 absorption saturates when the far-IR continuum approaches
the optically thick regime ($\tau_{100}>0.5$), as the far-IR field also
saturates and only the externalmost shells of the source contribute to the
absorption. The 
location of the data points in Fig.~\ref{models}a-b suggests that this
optically thick regime for both the continuum and the OH65 transition
applies to the majority of sources with
$W_{\mathrm{eq}}(\mathrm{OH65})\gtrsim20$ \kms. In contrast,
sources like the prototypical starbursts M~82 and
  NGC~253, the HII galaxy UGC~5101, and the AGNs NGC~1068 and NGC~6240 have
not developed these optically thick, prominent far-IR emitting 
structures\footnote{The Compton-thick AGNs
  NGC~1068 and NGC~6240 \citep[e.g.][]{bur11} are weak in OH65, probably
  because the X-ray absorber in these sources is too compact and hot to
  contribute significantly to the far-IR emission.}, though some of them
   may contain an embryo\footnote{In the deeply buried, edge-on
starburst+AGN NGC~4945 \citep[e.g.][]{spo00}, there is most likely an
optically thick far-IR emitting region, but either it is too cold to generate
strong OH65 or the continuum emission from that component is highly
diluted.}.
 
The deficit in [C {\sc ii}] is produced because
the cooling line cannot track the increase of $T_{\mathrm{dust}}(G_0)$
\citep[e.g.][]{kau99}. 
The region where carbon is ionized, and thus the total number of emitting
C$^+$ ions, is restricted to $\tau_{\mathrm{FUV}}\sim1$ regardless of $G_0$,
but the increase of $G_0$ increases the FIR continuum
-thus lowering [C {\sc ii}]/FIR.
Nevertheless, the presence of a [C {\sc ii}] deficit is not critically
sensitive to the physical details of the PDRs and can be mostly understood in
terms of the global $L_{\mathrm{FIR}}/M_{\mathrm{H2}}$: a strong 
upper limit on the [C {\sc ii}] emission 
is given by (G-A08) 
\begin{equation}
\frac{L_{\mathrm{[C\,{\sc II}]}}}{L_{\mathrm{FIR}}}<8.3\times10^{-4}\times\left(\frac{X_{\mathrm{C^+}}}{4\times10^{-5}}\right)\times\left(\frac{300\,L_{\odot}/M_{\odot}}{L_{\mathrm{FIR}}/M_{\mathrm{H2}}}\right), 
\label{eq:cplus}
\end{equation}
which assumes optically thin [C {\sc ii}] emission with
$T_{\mathrm{ex}}>>E_{\mathrm{upper}}=91$ K and up to
$1/3$ of all carbon ionized (for solar metallicities), which is high for
molecular regions. In the optically thick regime, line saturation and
extinction effects maintain a low [C {\sc ii}]/FIR even in case of high
$M_{\mathrm{H2}}$.  
Equation~(\ref{eq:cplus}) directly links the [C {\sc ii}]/FIR
ratio with $L_{\mathrm{FIR}}/M_{\mathrm{H2}}$, and shows that strong 
[C {\sc ii}] deficits are unavoidable in galaxies with high
$L_{\mathrm{FIR}}/M_{\mathrm{H2}}$, i.e. those that are also strong in OH65 
(Figs.~\ref{models}a-b).

The single component models, however, cannot account for the observed
  $f25/f60$ with the \tdust\ and $\tau_{100}$ inferred from the other panels of
  Fig.~\ref{models}. Indeed, 
the main consequence of assuming coexistent OH65 absorption and [C {\sc ii}]
emission, i.e. modeling only one OH transition with a fixed covering
factor of unity, is the underestimation of \tdust\ and $\tau_{100}$
(e.g. see multitransition, composite models for NGC~4418, Arp~220, and Mrk~231
in G-A12 and G-A14). Extended and optically thin regions of (U)LIRGs are
emitters of both [C {\sc ii}] and 65 $\mu$m continuum \citep{dia14}, 
and will dilute both the circumnuclear OH65 absorption and the 
[C {\sc ii}] deficit.

\begin{figure*}
\includegraphics[angle=0,scale=.65]{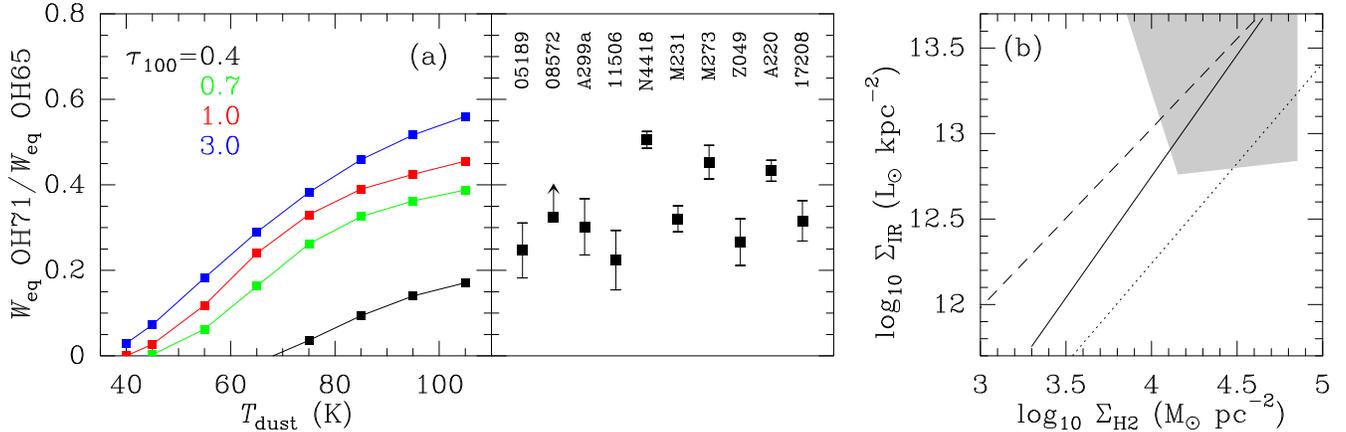}
\caption{
 a) Modeled $W_{\mathrm{eq}}(\mathrm{OH71})/W_{\mathrm{eq}}(\mathrm{OH65})$
as a function of \tdust\ for $\tau_{100}\ge0.4$, together with the 
ratios measured for the 10 sources where OH71 is detected, indicating
$T_{\mathrm{dust}}\gtrsim60$ K. 
b) Model-derived $L_{\mathrm{IR}}$ versus H$_2$ mass in the surface
density plane. The grey parallelogram identifies the region favored 
by the composite models ($C_{\mathrm{thick}}$, \S\ref{compc}) with
$\tau_{100}=0.5-5$ and $T_{\mathrm{dust}}\ge60$ K. Solid, dashed, and dotted
lines show the fits to (U)LIRGs/mergers/SMGs given in \cite{dad10},
\cite{gbu12}, and \cite{gen10}, respectively.}
\label{fig4}
\end{figure*}

\subsection{The OH71/OH65 ratio and Composite models} 
\label{compc}  

The ratio of the OH71 to the OH65 doublet absorption enables a better
  estimate of \tdust\ and $\tau_{100}$, independent of the covering factor of
  the continuum by the excited OH. The modeled ratio of the total doublet
  equivalent widths,
$W_{\mathrm{eq}}(\mathrm{OH71})/W_{\mathrm{eq}}(\mathrm{OH65})$, 
is plotted as a function of \tdust\ (for $\tau_{100}\ge0.4$) in
Fig.~\ref{fig4}a, and compared with the measured ratios. The observed
minimum $W_{\mathrm{eq}}(\mathrm{OH71})/W_{\mathrm{eq}}(\mathrm{OH65})\gtrsim0.2$
ratio indicates $T_{\mathrm{dust}}\ge60$ K, significantly higher than
  inferred in \S\ref{singlec}. This lower limit is still very conservative for
  hot {\em sub}components in some sources, where higher-lying transitions of
  OH (at 53.0 and 56 $\mu$m) and \hdo\ indicate $T_{\mathrm{dust}}\gtrsim90$ K
  (G-A12; G-A14; Falstad et al., in preparation). In addition, unless the dust
  is very warm, the measured ratios in Fig.~\ref{fig4}a favor
  $\tau_{100}\gtrsim0.7$, equivalent to $N_{\mathrm{H}}\gtrsim10^{24}$ \cmd. 

To account for the trends in Fig.~\ref{correl} together with the high
  \tdust\ inferred from
  $W_{\mathrm{eq}}(\mathrm{OH71})/W_{\mathrm{eq}}(\mathrm{OH65})$ in many of
  the sources, we relax the assumptions in \S\ref{singlec} and use composite
  models. It is assumed 
here that the bulk of the sources can be described by an optically
thick very warm component ($C_{\mathrm{thick}}$) that accounts for the OH65
absorption, and a colder, optically thin (presumably more extended) component
($C_{\mathrm{thin}}$)
that accounts for the bulk of the [C {\sc ii}] emission but gives negligible
OH65 absorption. The physical conditions are: $(i)$ $C_{\mathrm{thick}}$: two
models with $T_{\mathrm{dust}}=95$ and $60$ K are considered (hereafter $M_1$
and $M_2$, respectively); in both, a nominal $\tau_{100}=1$ is
used. $(ii)$ $C_{\mathrm{thin}}$: this is itself a model composed of two
optically thin dust components with $T_{\mathrm{dust}}=65$ and $30$ K,
consistent with the $G_0$-range derived by \cite{far13} and with results
independent of $\tau_{100}$ for values $\lesssim0.1$. The mix of these two
components is governed by 
$\beta_{65-30}=L_{\mathrm{IR}}^{65}/(L_{\mathrm{IR}}^{65}+L_{\mathrm{IR}}^{30})$,
i.e. the fraction of the optically thin IR luminosity arising from the warm
65 K component, which is fixed to $0.75$ and $0.25$ for $M_1$ and $M_2$,
respectively. The values of \tdust\ and $\beta_{65-30}$ for
  $C_{\mathrm{thin}}$ are chosen to bracket the observed $f25/f60$
  and $f60/f100$ colors for sources that have weak -but measurable- OH65
  absorption. 
We adopt here a typical $n_{\mathrm{H}}=10^3$ \cmt\ to simulate the [C
{\sc ii}] emission \citep[e.g.][]{mal01,par13}. 
The mix of $C_{\mathrm{thick}}$ and $C_{\mathrm{thin}}$ is described by the
only free parameter 
$\beta_{\mathrm{thick}}=L_{\mathrm{IR}}^{\mathrm{thick}}/L_{\mathrm{IR}}^{\mathrm{total}}$,
the fraction of the total IR luminosity arising from the $C_{\mathrm{thick}}$ 
component. 
  
Results for these models are shown in Fig.~\ref{models};
blue/green curves and circles correspond to models $M_1/M_2$, respectively. 
The circles on these curves represent a sequence of models where
$\beta_{\mathrm{thick}}$ is varied from 0 to 1 in intervals of $0.1$, with
$\beta_{\mathrm{thick}}=1$ indicated with a red circle. According to this
approach, the source position within the different observational planes is
interpreted in terms of the overall energetic relevance of the optically thick,
very warm structure ($\beta_{\mathrm{thick}}$). For sources with 
$W_{\mathrm{eq}}(\mathrm{OH65})\lesssim6$ \kms, 
$\beta_{\mathrm{thick}}\lesssim10$\%, while sources with 
$W_{\mathrm{eq}}(\mathrm{OH65})\gtrsim20$ \kms\ are characterized by
$\beta_{\mathrm{thick}}\gtrsim30-50$\% for $M_1-M_2$, respectively.
In the latter objects, a fraction of the optically thin emission is
  expected to be reemission by dust heated by the optically thick component
  \citep{soi99,gon04}, and thus $C_{\mathrm{thick}}$ most likely dominates the
  output of these galaxies. 
The modeled curves are consistent with the steep increase of
$W_{\mathrm{eq}}(\mathrm{OH65})$ with decreasing [C {\sc ii}]/FIR below
$\sim10^{-3}$, and with increasing $L_{\mathrm{FIR}}/M_{\mathrm{H2}}$ above
$\approx100$ \Lsun/Msun. The [C {\sc ii}] emission is still
underpredicted in some sources, which may suggest significant contributions by
(diffuse) ionized gas.

\section{Conclusions}

Absorption in high-lying transitions of
  molecules with high dipolar moment and level spacing (i.e. mostly light
  hydrides), represented by OH65 and OH71, has been shown here to be strong in
  most local ULIRGs ($79$\%) and in several LIRGs. Despite the high 
  columns inferred in galaxies with high $W_{\mathrm{eq}}(\mathrm{OH})$, their
  low [C {\sc ii}]/FIR and low $L'_{\mathrm{CO}}/L_{\mathrm{FIR}}$ suggest that
  both are associated with a ``deficit'' in $M_{\mathrm{H2}}$ relative to the
  far-IR continuum emission, accompanied by additional effects such as
  significant optical depth in the [C {\sc ii}] line and high excitation of CO. 
  High columns and \tdust\ but low $M_{\mathrm{H2}}/L_{\mathrm{FIR}}$ are
  indicative of high radiation densities and small volumes, with the high
  columns of gas and dust confined to small regions\footnote{The 
  effective size of the OH65 source can be determined by comparing the absolute
  $L_{\mathrm{IR}}$ (an upper limit of which is the observed luminosity) with
  its calculated value,  
  $4\pi\int\int B_{\nu}(T_{\mathrm{dust}})\,(1-\exp\{-\tau_{\nu}\})\,dS\,d\nu$,
  where $dS$ stands for the projected surface and $\tau_{\nu}$ is the
  continuum optical depth at frequency $\nu$ along the corresponding line of
  sight. For reference, the effective radius is $R_{\mathrm{eff}}\approx170$
  pc for $L_{\mathrm{IR}}=10^{12}$ \Lsun\ ($R_{\mathrm{eff}}\propto
  L_{\mathrm{IR}}^{1/2}$), $T_{\mathrm{dust}}=70$ K, and 
  $\tau_{100}=1$, which is a lower limit to the physical size in some galaxies
  owing to clumpiness. The diagnostics in Fig.~\ref{correl} and
  model results in Fig.~\ref{models} are nevertheless
  independent of sizes and adopted distances.\label{foot:size}} 
  around the bright, buried 
  illuminating source(s) (nearly) dominating the galaxy output. The
  relationship between the model parameterization used here and that in terms
  of the dominant exciting source (AGN or starburst), volume and column
    densities, and ionization parameter \citep[][G-C11]{abe09,fis14}, as
    well as the origin of the deficit in fine-structure lines other than [C
    {\sc ii}], will be explored in future work.

The inferred column densities associated with
$W_{\mathrm{eq}}(\mathrm{OH65})\gtrsim20$ \kms, 
$N_{\mathrm{H}}\gtrsim10^{24}$ \cmd, are higher than those
derived from the silicate strength at $9.7$ $\mu$m. 
Models by \cite{sir08} indicate that the observed $\tau_{9.7}\lesssim4$
  can be explained with $\tau_{\mathrm{V}}\lesssim300$ 
  ($N_{\mathrm{H}}\lesssim5\times10^{23}$ \cmd). 
Since the OH65 regions/structure will block all the inner mid-IR
emission passing through it, the observed mid-IR emission and associated
silicate absorption are biased toward relatively unabsorbed mid-IR emitting
regions. Likewise, several sources in the sample show mid-IR AGN signatures  
as [Ne {\sc v}] emission \citep[IRAS~05189-2524 and
Mrk~273;][V09]{arm07} or an optical Broad Line Region (e.g. Mrk~231), and
our direct view of this emission indicates tiny absorbing columns in
comparison with those inferred from OH65. If the OH65 absorption is
generated in a circumnuclear disk/torus/cocoon, either the combination of
scale height/inclination, and/or clumpiness are required to account for the
apparent decrease of extinction with decreasing wavelength. 
In sources with high contrast in mid-to-far infrared extinction,
either extreme clumpiness or important inclination effects are necessarily
involved. In other sources, extinction of the mid-IR emission by the OH65
structure is consistent with their low $f25/f60$ ratio.   
Though these structures are variably clumpy, the OH65 bimodality 
(Fig.~\ref{distrib}b-c) suggests that they are coherent and quickly formed, 
and provide an effective way to obscure the signposts of AGNs at shorter
wavelengths in some sources.

In the direction of the warm, optically thick regions/structure where
the OH65 absorption is produced, the surface density of both $L_{\mathrm{IR}}$
and H$_{2}$ mass are significantly higher than the
average values previously estimated for the areas within
the half-light radius (from CO or optical/UV). 
The grey parallelogram in Fig.~\ref{fig4}b indicates the location 
of the $C_{\mathrm{thick}}$ component for the ten sources where
both OH65 and OH71 are detected:
$\Sigma_{\mathrm{H2}}\sim10^{3.8}-10^{4.7}$ \Msun\ pc$^{-2}$
($\tau_{100}\sim0.5-5$) and $\Sigma_{\mathrm{IR}}\gtrsim10^{12.8}$  
\Lsun\ kpc$^{-2}$ ($T_{\mathrm{dust}}\ge60$ K).  
The latter high fluxes are consistent with previous estimates
  using sizes derived from radio emission and strengthen the role of
  radiation pressure support \citep[][see their
  Fig.~3]{sco04,tho05}. For starburst-dominated sources and using a
\cite{cha03} IMF, the corresponding SFRs are
$\Sigma_{\mathrm{SFR}}\gtrsim10^{2.8}$ \Msun\ yr$^{-1}$
kpc$^{-2}$. For hot {\em sub}components in some sources like the
$C_{\mathrm{core}}$ component of NGC~4418,
$\Sigma_{\mathrm{IR}}\gtrsim10^{14}$ \Lsun\ kpc$^{-2}$ 
and $\tau_{100}\gtrsim8$ ($N_{\mathrm{H}}\gtrsim10^{25}$ \cmd)
on spatial scales of $\approx20$ pc 
\citep[G-A12,][]{sak13,cos13,var14}.  

The solid, dashed, and dotted black lines in Fig.~\ref{fig4}b are
extrapolations of the fits found in previous studies for subsamples of
  (U)LIRGs/mergers/SMGs. A pure SF scenario involves the
shadowing of $\gtrsim70$ (for $\tau_{100}\gtrsim1$) star-forming regions, each
with $A_V\sim10$ mag and $G_0\gtrsim10^4$, on spatial scales of 
$\mathrm{a\,few}\times(10-100)$ pc; on spatial scales of a few parsecs,
  there is apparently no analog star-forming region close to the Galactic 
center.\footnote{In Sgr~B2(M), the peak of one of the most active and
  optically thick molecular cloud complexes in the Milky Way,
    $W_{\mathrm{eq}}(\mathrm{OH65})\approx2$ 
  \kms\ as measured with {\it ISO}/FP \citep{pol07}, consistent with its
  moderate effective $T_{\mathrm{dust}}\sim34$ K, 
  $\Sigma_{\mathrm{IR}}\sim10^{11.8}$ \Lsun\ kpc$^{-2}$, and
  $L_{\mathrm{FIR}}/M_{\mathrm{H2}}\sim22$ \Lsun/\Msun\ for a gas-to-dust
  ratio by mass of 100 \citep{etx13}. OH65 is not detected towards 
  Sgr~A$^*$ and its circumnuclear disk \citep{goi13}, and is detected in
  emission towards the Orion bar PDR indicating collisional excitation 
  in warm and dense gas \citep{goi11}.} The
implied gas consumption timescales are
$\tau_{\mathrm{gas}}=(25-3.5)\times\tau_{100}$ Myr 
for $T_{\mathrm{dust}}=60-95$ K, comparable to
those estimated for extreme sources exhibiting powerful 
OH outflows driven by buried AGNs \citep{stu11}.
On the other hand, 21 sources in our sample were analyzed in the OH 119 $\mu$m
transition by V13, and 12 (9 with
$W_{\mathrm{eq}}(\mathrm{OH65})>10$ \kms) were found to have
$|v_{84}|\gtrsim500$ \kms\ (the velocity below which 84\% of the absorption
takes place), most likely indicative of significant AGN feedback. The
OH65-(U)LIRG phase may thus represent the starburst-AGN 
co-evolution phase in its shortlived most buried/active stage.

\acknowledgments

PACS has been developed by a consortium of institutes
led by MPE (Germany) and including UVIE (Austria); KU Leuven, CSL,
IMEC (Belgium); CEA, LAM (France); MPIA (Germany);  
INAFIFSI/OAA/OAP/OAT, LENS, SISSA (Italy); IAC (Spain). This development
has been supported by the funding agencies BMVIT (Austria),
ESA-PRODEX (Belgium), CEA/CNES (France), DLR (Germany), ASI/INAF (Italy), and 
CICYT/MCYT (Spain). E.G-A is a Research Associate at the Harvard-Smithsonian
CfA, and thanks the Spanish Ministerio de Econom\'{\i}a y Competitividad for
support under projects AYA2010-21697-C05-0 and FIS2012-39162-C06-01. 
E.G-A and H.A.S. acknowledge partial support from NHSC/JPL RSA 1455432;
H.A.S acknowledges NASA grant NNX14AJ61G. Basic
research in IR astronomy at NRL is funded by the US-ONR;  
J.F. acknowledges support from NHSC/JPL subcontracts 139807 and 1456609.
S.V. and M.M. acknowledge partial support from NHSC/JPL RSA
1427277 and 1454738.
This research has made use of NASA's Astrophysics Data System
and of GILDAS (http://www.iram.fr/IRAMFR/GILDAS).

{\it Facilities:} \facility{{\it Herschel Space Observatory} (PACS)}.


\begin{thebibliography}{}

  \bibitem[Aalto et al.(1997)]{aal97} Aalto, S., Radford, S. J. E., Scoville,
    N. Z., \& Sargent, A. I. 1997, \apj, 475, L107
  \bibitem[Abel et al.(2009)]{abe09} Abel, N. P., Dudley, C., Fischer, J.,
    Satyapal, S., \& van Hoof, P. A. M. 2009, \apj, 701, 1147
  \bibitem[Armus et al.(2007)]{arm07} Armus, L., Charmandaris, V.,
    Bernard-Salas, J., et al. 2007, \apj, 656, 148
  \bibitem[Brauher et al.(2008)]{bra08} Brauher, J. R., Dale, D. A., \& Helou,
    G. 2008, \apjs, 178, 280
  \bibitem[Burlon et al.(2011)]{bur11} Burlon, D., Ajello, M., Greiner, J.,
    Comastri, A., Merloni, A., \& Gehrels, N. \apj, 728, 58
  \bibitem[Casoli et al.(1999)]{cas99} Casoli, F., Willaime, M.-C.,
    Viallefond, F., \& Gerin, M. 1999, A\&A, 346, 663
  \bibitem[Chabrier(2003)]{cha03} Chabrier, G. 2003, \apj, 586, L133
  \bibitem[Chung et al.(2009)]{chu09} Chung, A., Narayanan, G., Yun, M. S.,
    Heyer, M., \& Erickson, N. R. 2009, AJ, 138, 858
  \bibitem[Costagliola et al.(2013)]{cos13} Costagliola, F., Aalto, S.,
    Sakamoto, K., Mart\'{\i}n, S., Beswick, R., Muller, S., \& Kl\"ockner,
    H.-R. 2013, A\&A, 556, A66
  \bibitem[Daddi et al.(2010)]{dad10} Daddi, E., Elbaz, D., Walter, F., et al.
    2010, \apj, 714, L118
  \bibitem[D\'{\i}az-Santos et al.(2013)]{dia13} D\'{\i}az-Santos, T., Armus,
    L., Charmandaris, V., et al. 2013, \apj, 774, 68
  \bibitem[D\'{\i}az-Santos et al.(2014)]{dia14} D\'{\i}az-Santos, T., Armus,
    L., Charmandaris, V., et al. 2014, \apj, in press (arXiv:1405.3983)
  \bibitem[Downes et al.(1993)]{dow93} Downes, D., Solomon, P. M., \& Radford,
    S. J. E. 1993, \apj, 414, L13
  \bibitem[Etxaluze et al.(2013)]{etx13} Etxaluze, M., Goicoechea, J. R.,
    Cernicharo, J., Polehampton, E. T., Noriega-Crespo, A., Molinari, S.,
    Swinyard, B. M., Wu, R., \& Bally, J. 2013, A\&A, 556, A137
  \bibitem[Farrah et al.(2013)]{far13} Farrah, D.; Lebouteiller, V.; Spoon,
    H. W. W., et al. 2013, \apj, 776, 38 
  \bibitem[Fischer et al.(1999)]{fis99} Fischer, J., Luhman, M. L., Satyapal,
    S., et al. 1999, \apss, 266, 91  
  \bibitem[Fischer et al.(2010)]{fis10} Fischer, J., Sturm, E.,
    Gonz\'alez-Alfonso, et al. 2010, A\&A, 518, L41
  \bibitem[Fischer et al.(2014)]{fis14} Fischer, J., Abel, N. P.,
    Gonz\'alez-Alfonso, E., Dudley, C. C., Satyapal, S., \& van Hoof,
    P. A. M. 2014, ApJ, in press (arXiv:1409.2521) 
  \bibitem[Garc\'{\i}a-Mar\'{\i}n et al.(2006)]{gar06} Garc\'{\i}a-Mar\'{\i}n, 
    M., Colina, L., Arribas, S., Alonso-Herrero, A., \& Mediavilla, E. 2006, 
    \apj, 650, 850
  \bibitem[Garc\'{\i}a-Burillo et al.(2012)]{gbu12} Garc\'{\i}a-Burillo, S.,
    Usero, A., Alonso-Herrero, A., Graci\'a-Carpio, J., Pereira-Santaella, M.,
    Colina, L., Planesas, P., \& Arribas, S. 2012, A\&A, 539, A8
  \bibitem[Genzel et al.(2010)]{gen10} Genzel, R., Tacconi, L. J.,
    Graci\'a-Carpio, J. et al. 2010, MNRAS, 407, 2091 
  \bibitem[Goicoechea \& Cernicharo(2002)]{goi02} Goicoechea, J. R. \&
    Cernicharo, J. 2002, \apj, 576, L77
  \bibitem[Goicoechea et al.(2011)]{goi11} Goicoechea, J. R., Joblin, C.,
    Contursi, A., Bern\'e, O., Cernicharo, J., Gerin, M., Le Bourlot, J.,
    Bergin, E. A., Bell, T. A., \& R\"ollig, M. 2011, A\&A, 530, L16
  \bibitem[Goicoechea et al.(2013)]{goi13} Goicoechea, J. R., Etxaluze, M.,
    Cernicharo, J., et al. 2013, \apj, 769, L13
  \bibitem[Gonz\'alez-Alfonso et al.(2004)]{gon04} Gonz\'alez-Alfonso,
    E., Smith, H. A., Fischer, J., \& Cernicharo, J. 2004, ApJ, 613, 247
  \bibitem[Gonz\'alez-Alfonso et al.(2008)]{gon08} Gonz\'alez-Alfonso,
    E., Smith, H. A., Ashby, M. L. N., Fischer, J., Spinoglio, L., \& Grundy,
    T. W. 2008, ApJ, 675, 303 (G-A08) 
  \bibitem[Gonz\'alez-Alfonso et al.(2012)]{gon12} Gonz\'alez-Alfonso,
    E., Fischer, J., Graci\'a-Carpio, J., et al. 2012, A\&A, 541, A4 (G-A12)
  \bibitem[Gonz\'alez-Alfonso et al.(2013)]{gon13} Gonz\'alez-Alfonso,
    E., Fischer, J., Bruderer, S., et al. 2013, A\&A, 550, A25
  \bibitem[Gonz\'alez-Alfonso et al.(2014a)]{gon14} Gonz\'alez-Alfonso,
    E., Fischer, J., Graci\'a-Carpio, J., et al. 2014, A\&A, 561, A27 (G-A14)
  \bibitem[Gonz\'alez-Alfonso et al.(2014b)]{gon14b} Gonz\'alez-Alfonso,
    E., Fischer, J., Aalto, S., \& Falstad, N. 2014, A\&A, 567, A91
  \bibitem[Goulding et al.(2012)]{gou12} Goulding, A. D., Alexander, D. M.,
    Bauer, F. E., Forman, W. R., Hickox, R. C., Jones, C., Mullaney, J. R., \&
    Trichas, M. 2012, \apj, 755, 5
  \bibitem[Graci\'a-Carpio et al.(2011)]{gra11} Graci\'a-Carpio, J., Sturm,
    E., Hailey-Dunsheath, S., et al. 2011, ApJ, 728, L7 (G-C11)
  \bibitem[Helou et al.(1988)]{hel88} Helou, G., Khan, I. R., Malek, L., \&
    Boehmer, L. 1988, ApJS, 68, 151
  \bibitem[Henkel et al.(1994)]{hen94} Henkel, C., Whiteoak, J. B., \&
    Mauersberger, R. 1994, A\&A, 284, 17
  \bibitem[Hollenbach et al.(1991)]{hol91} Hollenbach, D. J., Takahashi, T.,
    \& Tielens, A. G. G. M. 1991, \apj, 377, 192
  \bibitem[Houghton et al.(1997)]{hou97} Houghton, S., Whiteoak, J. B.,
    Koribalski, B., Booth, R., Wiklind, T., \& Wielebinski, R. 1997, A\&A,
    335, 923
  \bibitem[Kaufman et al.(1999)]{kau99} Kaufman, M. J., Wolfire, M. G.,
    Hollenbach, D. J., \&  Luhman, M. L. 1999, \apj, 527, 795
  \bibitem[Kim et al.(1998)]{kim98} Kim, D.-C., Veilleux, S., \& Sanders,
    D. B. 1998, \apj, 508, 627
  \bibitem[Luhman et al.(1998)]{luh98} Luhman, M. L., Satyapal, S., Fischer,
    J., Wolfire, M. G., Cox, P., Lord, S. D., Smith, H. A., Stacey, G. J., \&
    Unger, S. J. 1998, \apj, 504, L11
  \bibitem[Luhman et al.(2003)]{luh03} Luhman, M. L., Satyapal, S., Fischer,
    J., Wolfire, M. G., Sturm, E., Dudley, C. C., Lutz, D., \& Genzel,
    R. 2003, \apj, 594, 758
  \bibitem[Malhotra et al.(2001)]{mal01} Malhotra, S., Kaufman, M. J.,
    Hollenbach, D., et al. 2001, \apj, 561, 766
  \bibitem[Meijerink et al.(2011)]{mei11} Meijerink, R., Spaans, M., Loenen,
    A. F., \& van der Werf, P. P. 2011, A\&A, 525, A119
  \bibitem[Mirabel et al.(1990)]{mir90} Mirabel, I. F., Booth, R. S.,
    Johansson, L. E. B., Garay, G., \& Sanders, D. B. 1990, A\&A, 236, 327
  \bibitem[Narayanan et al.(2005)]{nar05}Narayanan, D., Groppi, C. E.; Kulesa,
    C. A., \& Walker, C. K. 2005, \apj, 630, 269
  \bibitem[Papadopoulos et al.(2012)]{pap12} Papadopoulos, P. P., van der
    Werf, P. P., Xilouris, E. M., Isaak, K. G., Gao, Y., \& M\"uhle, S. 2012,
    MNRAS, 426, 2601 
  \bibitem[Parkin et al.(2013)]{par13} Parkin, T. J., Wilson, C. D., Schirm,
    M. R. P., et al. 2013, \apj, 776, 65
  \bibitem[Pilbratt et al.(2010)]{pil10} Pilbratt, G. L.; Riedinger, J. R.;
    Passvogel, T., et al. 2010, A\&A, 518, L1
  \bibitem[Poglitsch et al.(2010)]{pog10} Poglitsch, A., Waelkens, C., Geis,
    N., et al. 2010, A\&A, 518, L2
  \bibitem[Polehampton et al.(2007)]{pol07} Polehampton, E. T., Baluteau, J.-P.,
    Swinyard, B. M., Goicoechea, J. R., Brown, J. M., White, G. J.,
    Cernicharo, J., \& Grundy, T. W. 2007, MNRAS, 377, 1122
 \bibitem[Rupke et al.(2005)]{rup05} Rupke, D. S., Veilleux, S, \& Sanders,
   D. B. 2005, \apjs, 160, 87
 \bibitem[Sakamoto et al.(2013)]{sak13} Sakamoto, K., Aalto, S., Costagliola,
   F., Mart\'{\i}n, S., Ohyama, Y., Wiedner, M. C.; Wilner, D. J. 2013, \apj,
   764, 42
 \bibitem[Sanders et al.(2003)]{san03} Sanders, D. B., Mazzarella, J. M.,
    Kim, D.-C., Surace, J. A., \& Soifer, B. T. 2003, \aj, 126, 1607 
  \bibitem[Scoville et al.(1983)]{sco83} Scoville, N. Z., Young, J. S., \&
    Lucy, L. B. 1983, \apj, 270, 443
  \bibitem[Scoville(2004)]{sco04} Scoville, N. Z. 2004, in The Neutral ISM in
    Starburst Galaxies, ed. S. Aalto, S. H\"uttemeister, \& A. Pedlar, ASP
    Conf. Ser., 320, 253 
  \bibitem[Sirocky et al.(2008)]{sir08} Sirocky, M. M., Levenson, N. A.,
    Elitzur, M., Spoon, H. W. W., \& Armus, L. 2008, \apj, 678, 729
  \bibitem[Soifer et al.(1999)]{soi99} Soifer, B. T., Neugebauer, G.,
    Matthews, K., Becklin, E. E., Ressler, M., Werner, M. W., Weinberger,
    A. J., \& Egami, E. 1999, \apj, 513, 207 
  \bibitem[Solomon et al.(1997)]{sol97} Solomon, P. M., Downes, D., Radford,
    S. J. E., \& Barrett, J. W. 1997, \apj, 478, 144
  \bibitem[Spoon et al.(2000)]{spo00} Spoon, H. W. W., Koornneef, J.,
    Moorwood, A. F. M., Lutz, D., \& Tielens, A. G. G. M. 2000, A\&A, 357, 898
  \bibitem[Spoon et al.(2007)]{spo07} Spoon, H. W. W., Marshall, J. A., Houck,
    J. R., Elitzur, M., Hao, L., Armus, L., Brandl, B. R., \& Charmandaris,
    V. 2007, \apj, 654, L49
  \bibitem[Spoon et al.(2013)]{spo13} Spoon, H. W. W.,  Farrah, D.,
  Lebouteiller, V., et al. 2013, ApJ, 775, 127 
  \bibitem[Sturm et al.(2011)]{stu11} Sturm, E., Gonz\'alez-Alfonso, E.,
    Veilleux, S., et al. 2011, ApJ, 733, L16 
  \bibitem[Surace et al.(2004)]{sur04} Surace, J. A., Sanders, D. B., \&
    Mazzarella, J. M. 2004, AJ, 127, 3235
  \bibitem[Thompson et al.(2005)]{tho05} Thompson, T. A., Quataert, E., \&
    Murray, N. 2005, \apj, 630, 167
  \bibitem[Tielens \& Hollenbach(1985)]{tie85} Tielens, A. G. G. M. \&
    Hollenbach, D. 1985, \apj, 291, 722
  \bibitem[Varenius et al.(2014)]{var14} Varenius, E., Conway, J. E.,
    Mart\'{\i}-Vidal, I., Aalto, S., Beswick, R., Costagliola, F., \&
    Kl\"ockner, H.-R. 2014, A\&A, 566, A15
  \bibitem[Veilleux et al.(1995)]{vei95} Veilleux, S., Kim, D.-C., Sanders,
    D. B., Mazzarella, J. M., \& Soifer, B. T. 1995, \apjs, 98, 171
  \bibitem[Veilleux et al.(1999)]{vei99} Veilleux, S., Kim, D.-C., \& Sanders,
    D. B. 1999, \apj, 522, 113
  \bibitem[Veilleux et al.(2009)]{vei09} Veilleux, S., Rupke, D. S. N., Kim,
    D.-C., et al. 2009, \apjs, 182, 628 (V09) 
  \bibitem[Veilleux et al.(2013)]{vei13} Veilleux, S., Mel\'endez, M.; Sturm,
    E., et al. 2013, ApJ, 776, 27 (V13)
  \bibitem[V\'eron-Cetty \& V\'eron(2006)]{ver06} V\'eron-Cetty,
    M.-P. \& V\'eron, P. 2006, A\&A, 455, 773 
  \bibitem[Wei\ss\ et al.(2005)]{wei05} Wei\ss, A., Walter, F., \& Scoville,
    N. Z. 2005, A\&A, 438, 533
\end{thebibliography}
\end{document}